# Calculation of Dark Matter as a Feature of Space–Time

## Two Geometries, Derivation of: Rotation Curves, Universal Dark Matter Density Formula and Tully–Fisher Relations

Peter H. Handel[1] · Klara E. Splett[1,2]

## Abstract
We derive the first analytical formula for the density of "Dark Matter" (DM) at all length scales, thus also for the rotation curves of stars in galaxies, for the baryonic Tully–Fisher relation and for planetary systems, from Einstein's equations (EE) and classical approximations, in agreement with observations. DM is defined in Part I as the energy of the coherent gravitational field of the universe, represented by the additional equivalent ordinary matter (OM), needed at all length scales, to explain classically, with inclusion of the OM, the *observed coherent* gravitational field. Our derivation uses both EE and the Newtonian approximation of EE in Part I, to describe semi-classically in Part II the advection of DM, created at the level of the universe, into galaxies and clusters thereof. This advection happens proportional with their own classically generated gravitational field g, due to self-interaction of the gravitational field. It is based on the universal formula $\rho_D = \lambda g g'^2$ for the density $\rho_D$ of DM advected into medium and lower scale structures of the observable universe, where $\lambda$ is a universal constant fixed by the Tully–Fisher relations. Here $g'$ is the gravitational field of the universe; $g'$ is in main part its own source, as implied in Part I from EE. We start from a simple electromagnetic analogy that helps to make the paper generally accessible. This paper allows for the first time the exact calculation of DM in galactic halos and at all levels in the universe, based on EE and Newtonian approximations, in agreement with observations.

**Keywords**  Dark matter · General relativity · Astrophysics · Cosmology · Interdisciplinary · 1/f noise

✉ Peter H. Handel
  handelp@umsystem.edu; phasenoisehandel@gmail.com

  Klara E. Splett
  erikasplett@yahoo.de

[1] Department of Physics and Astronomy and Center for Nanoscience, University of Missouri St. Louis, St. Louis, MO 63131, USA

[2] Present Address: Peter H. Handel's Research Team, Kirschallee 6, 06809 Petersroda, Germany

## 1 Introduction

The notion of dark matter (DM) was introduced by Zwicky in 1933 to explain additional gravitational forces that were noticed in the universe, in very large structures. In the ensuing century, hundreds of studies have verified the existence of DM in large cosmic structures, in galaxies and their cluster formations. In galaxies, DM causes constant rotation velocities of stars in the periphery, instead of a $r^{-1/2}$ decrease predicted by Newtonian dynamics. In clusters of galaxies DM is needed to explain their stability and the presence of considerable gravitational lensing effects. So far DM revealed itself only through its gravitational effects, including lensing.

According to the "Lambda Cold Dark Matter" (ΛCDM) standard cosmological model, about 26.5% of the mass of the universe is DM [1]. Ordinary matter, including gas, plasma, dust, stars, planetary systems, galaxies and clusters, represents about 4.9%. The rest is considered to be dark energy [1], of unknown origin, represented by the cosmological constant in Einstein's Equations (EE).

This paper presents, *in two Parts,* our investigation that identified DM as gravitational field energy of the universe, also solving the DM problem for the first time, based only on general relativity (GR) and its classical, Newtonian, approximation that is applied wherever possible. ***Part I*** introduces the notion of coherent field mass, evaluates it as the DM, thus defined in the observable universe, and explores it both classically and in GR in Sects. 2.2–2.6. It finds DM to be dominant over ordinary matter at the larger and largest scales.

Based on this dominance of DM, ***Part II*** anticipates some features and results of any exact, all-encompassing solution of Einstein's equations for the observable universe, including structures, fields, their interaction and motion on all length scales. It yields a simple universal gravitational field self-interaction formula, a key for calculating DM at all lower scales of the observable universe. Surprisingly, this simple field interaction formula, this "key" leads to a remarkable simple GR-based solution of the DM problem in galaxies and other structures in Sects. 3.1–3.7. *It correctly describes this situation, where DM becomes mainly its own source at the largest scale. Readers may focus directly on this second Part, reading first* Sects. 3.1–3.7, ***then*** Sects. 2.3–2.6. ***They may accept the presence of five times more DM based on GR, or almost that much, classically. They first may want to see how this approach can explain all DM GR effects with surprisingly simple, quasi-classical means, including the rotation curves of stars in galaxies, the baryonic Tully–Fisher relations, DM distribution in colliding clusters of galaxies, and the cosmic web. They also may want to see the derivation of the general distribution law of DM in structures of any size.*** At the end, these readers may eventually want to go back to Sects. 2.1–2.6, to understand intuitively why classical Newtonian calculations were applicable in the second part, to the Ansatz on dominant DM in the universe, as they were used successfully in detail in Sects. 2.1–2.6, and compared to GR at each step.

Following the actual history of this investigation since 1975, we start in Sect. 2.1 with an elementary, generally accessible analogy, by first introducing both, the electromagnetic-coherent kinetic, and mechanical-"incoherent," kinetic energy contributions from electric currents. In this analogy, they correspond to dark and ordinary matter, as we shall see, in the gravitational case. We then introduce the coherent kinetic gravitational field notion, to evaluate the amount of DM at different length scales in Sect. 2.2, showing that almost all DM is actually generated at the scale of the observable universe, not independently at lower scales. Based on the spirit of GR, we early introduce *in* Sect. 2.2 *our* "DM key," a simple formula for the local density ρ of DM, which is caused by, and contributes to, the curvature of space–time.

In Sect. 2.3 we present our first general relativistic derivation of the result N′$r_s$, starting from the spherically symmetric Schwarzschild solution of Einstein's equations. which verifies the classic Sect. 2.1, Our Ansatz in Eq. (2), is verified on the basis of GR in Sect. 2.4.

Sect. 2.5 is dedicated to a detailed classical calculation of the energy of the coherent gravitational field of the observable universe in static, spherical, symmetry and of its relation to the energy of ordinary matter, expressed in the parameter s″.

Sect. 2.6 uses the result of the lowest order static GR calculation in spherical symmetry of the coherent gravitational energy density $t_{00}$ in Appendix 2, to obtain a GR derivation of the parameter s″ in spherical symmetry.

The somewhat laborious Part I is important. Indeed, it shows how even low-order GR calculations improve the Newtonian approximations in general, increasing the agreement with observations within an order of magnitude. The similarity, established at the level of the observable universe in two symmetries between Newtonian and GR approaches, provides the basis for applying the Newtonian approximation to DM in Part II, and the basis for the Ansatz that describes the accretion of universal DM to the gravitational fields of lower-scale structures.

In Part II, Sect. 3.1 brings an investigation of expected features of any exact GR solution for the whole observable universe at the level of planetary systems, galaxies or clusters thereof, and of the whole universe. This shows how the large amount of mainly self-generated DM present at the level of the whole universe coherently inflates the gravitational field of lower-level structures, with accreted DM described by an Ansatz of third-order in the deviations $h_{\mu\nu}$ of the metric coefficients $g_{\mu\nu}$ from their Euclidian values.

In Sect. 3.2 we explain the rotation curves of stars in galaxies, based on the observed general proportionality (also expected from the nonlinearity of the GR field equations and self-sourcing of DM) of the DM density accreted into galaxies, with their own gravitational field, generated by their baryonic matter. This is done with our "DM key."

Sect. 3.3 derives the empirical Tully–Fisher relation relating the total ordinary (baryonic) mass of a spiral galaxy to its peripheric rotation speed; on this basis we adjust both the universal "key" for finding the DM accreted into lower-scale structures and the exact form of the rotation curves of stars in galaxies. We then look for possible DM effects in the solar system in Sect. 3.4, and find them negligible.

In Sect. 3.5 we describe qualitatively, based on GR, how the self-interaction of the coherent gravitational field of the universe *can yield the cosmic web*, with non-uniformities initially triggered by quantum fluctuations as we assume, and decorated by ordinary matter. Sect. 3.6 brings the connection with the present ΛCDM cosmological model, with references and a discussion of our first derivation of the well-known Tully–Fisher relations from the classical approximation of GR.

We discuss our results and further research avenues in Sect. 3.7. We also present there our ***methods*** based on our gravitational generalization s″ of the s-parameter of our universal engineering formulas of the ubiquitous fundamental 1/f noise, as a new aspect of quantum mechanics.

The First Appendix describes the electromagnetic classic analogy both in cylindric, and in static, spherically symmetric, symmetry. The Second Appendix brings the details of the static, spherically symmetric, GR calculation in lowest order of the deviations $h_{\mu\nu}$ of the metric coefficients from the Euclidian metric.

This paper is interdisciplinary in nature, and the references to well-known astrophysical facts are not optimal, but are rather cited as examples. Previous research on DM mainly tried to identify weakly interacting particles that could be accounting for DM.

## 2 Classical and Relativistic Calculation of Dark Matter in Two Symmetries for the Universe

### 2.1 Conventional and Coherent Kinetic Effects

Consider first a n-type semiconductor wire carrying a current J and a current of particles J/e. Per unit length of the wire, the *conventional* kinetic energy of the drift motion of the current carriers of mass *m* and drift velocity $v_d$ is $N'mv_d^2/2$. On the other hand, the *coherent* magnetic term, i.e., the magnetic addition to this kinetic energy is $LJ^2/2c^2 = Le^2v_d^2N'^2/2c^2$. Here L is the autoinduction coefficient per unit length. *Only the latter, coherent, term is quadratic in N′, the number of carriers per unit length along the current.* At low N′ the coherent (magnetic, $\sim N'^2$) term can be neglected, but at large N′, e.g., in a copper wire or heavily doped semiconductor, the first term, linear in N′, is negligible. *We show here how this carries over to the gravidynamic domain, and how it explains the nature of DM. Although this paper is in essence completely independent of our universal quantum theory of fundamental 1/f noise, we show how it grew out of some basic considerations that also led to our 1966 turbulence theory of 1/f noise and to the related subsequent discovery of the universal* Quantum 1/f Effects *(Q1/fE), practically important in most domains of high-tech engineering. This was immediately applied* [2] *also to GR, at the same time, in 1975.*

Indeed, in earlier papers [2, 3] we have discussed the conventional and coherent gravidynamic Q1/fE, as well as the proposed indirect observation of gravitons in macroscopic streams of matter. The two gravidynamic Q1/fE were derived and discussed in analogy with the corresponding QED effects. To explain the two forms of fundamental 1/f noise observed in electronoc and microelectronic devices, we also identified the cause of these Q1/fE and the energy types, or terms in the hamiltonian, of the current carriers in semiconductors, connected with both forms of the QED-Q1/fE. For the conventional Q1/fE this term is the mechanical kinetic energy $mv_d^2/2$ of the drift velocity $v_d$ of the carriers, an energy proportional to the number N′ of carriers per unit length of the current path. As for the much larger coherent Q1/fE, dominant in large devices or semiconductor samples, it is connected to the coherent, collective, magnetic form of kinetic energy, associated with the drift of the carriers. *This form is proportional to the square of N′, and is therefore considered "coherent"*. The ratio between these two forms of kinetic energy in a given current or flow of particles defines the parameter *s, which we introduced to predict for engineers the fraction of coherent Q1/fE, compared to the conventional Q1/fE in solid state devices*. Dropping the subscript d in the drift velocity $v_d$, we obtain in cylindrical symmetry for the length unit of a long straight current J = N′ev, with $B_{outside} = 2$ J/cr, with CGS Gauss nonrationalized units

$$s = \left[\int B^2 d\tau/8\pi\right] \Big/ \left[N'mv^2/2\right]$$
$$= \left\{[4J^2/8\pi c^2]\left[\int 2\pi r dr/r^2\right]\right\} \Big/ \{N'mv^2/2\} \quad (1)$$
$$= e^2 N'^2 v^2 [0.25 + \ln(R_0/a)] / \{c^2 N'mv^2/2\}$$
$$= 2N'(e^2/mc^2)[0.25 + \ln(R_0/a)] \approx 2N'r_e.$$

Here J is again the electric current of the particles of mass m and charge e, B its magnetic field, and $R_0$ is the radius of the electric circuit, a cut-off for the logarithmic divergence. We introduced the classical radius of the electron $r_e = e^2/mc^2 = 2.8 \cdot 10^{-13}$ cm. The magnetic energy inside the radius of the wire *a*, if included also, with uniform current density, contributes the term 0.25, simply added to the logarithm here (see Appendix 1). This detail was suppressed in the second line of Eq. (1). The element of volume per unit length in cylindrical symmetry is $d\tau = 2\pi r dr$. The logarithm is considered of an order of magnitude not too far from unity and the rectangular bracket with it is set $\approx 1$.

*It is interesting that the ratio of the mechanical and magnetic field terms of the kinetic energy (both proportional to $v^2$) is independent of the drift velocity v, and should be constant also in the limit of vanishing velocity v = 0. It is a mass ratio, independent of the inclination of the world line, or world corridor, of the hyperbolic angle to the time axis in four dimensions.* With increased doping, as the number of current carriers in a salami slice of thickness $r_e$ increases and becomes larger than 1/2, the rest mass (or rest energy) that must be used to calculate the total kinetic drift energy of the carriers gradually shifts from the particles to their collective, coherent, electromagnetic field. We do not call the magnetic coupling "non-minimal," to distinguish it from the coupling through the electric field in the rest frame. They are the same coupling, seen from different frames of reference, related through Lorentz transformation. (see Appendix 1).

The magnetic field B can be written as $\mathbf{B} = \mathbf{v} \times \mathbf{E}/c$, where E is the electric field 2N′e/r that would be present in cylindric symmetry from the moving current carriers in the semiconductor in the absence of the neutralizing lattice. It can be obtained from **E** by Lorentz transformation.

In GR we expect to find also an analogue $\mathbf{v} \times \mathbf{g}/c$ of the magnetic field vector **B**, similar to $\mathbf{v} \times \mathbf{E}/c$, as the kinetic coherent gravitational field generated by the axial motion of a straight, very long, column of particles of mass m, with N′ particles per unit length. In Sect. 2.3 we show how the results of this conjecture, based on analogy, are tested in GR. This kinetic field is expected to be present in addition to the static gravitational field $g = -2GmN'/r$ (Gauss's law) in cylindrical symmetry. This coherent kinetic gravitational field is obtained from **g** by cross product with its drift (ordered, collective, component of the) velocity **v**. This is not a relativistic invariant calculation. It is done for low v, close to the local frame. We obtain thus for the gravidynamics case, by analogy,

$$\begin{aligned} s'' = s_g &= \left[ \int (\mathbf{v} \times \mathbf{g}/c)^2 d\tau / 8\pi G \right] \bigg/ \left[ N'mv^2/2 \right] \\ &= \left[ 4G m^2 N'^2 v^2 / (8\pi c^2 N' m v^2 / 2) \right] \int 2\pi r dr / r^2 \\ &= 2(GN'm/c^2)\left[0.25 + \ln(R_0/a)\right] \\ &\approx 2N'Gm/c^2 = N'r_s. \end{aligned} \quad (2)$$

Here we introduced the Schwarzschild radius $r_s = 2Gm/c^2$ of the particles, bodies, or systems of mass m in the beam, stream, or jet of matter, distributed with circular cross sectional area of approximate radius $a$. $R_0$ is again the radius of the very large circuit that would close the stream into a torus for cut-off. The gravitational energy inside the radius of the wire $a$, if included also, with uniform matter current density, contributes again the term 0.25, simply added to the logarithm here (see Appendix 1).

*The ratio s" is again independent of* $v_.$ It is again independent of the inclination of the world line, or world corridor, of the flowing matter, to the time axis. *Most important, the coherent gravitational energy term in the total kinetic energy is again proportional with $N'^2$, similar to the electrodynamic case.* (It was always known that in the rest frame, the gravitational energy per unit length of the column $-GN'^2m^2/4$ was coherent, i.e., also proportional to $N'^2$; see Appendix 1). In addition, there is also a contribution proportional to $v^2$ in a moving frame, like the mechanical kinetic energy, at low speed. *This corresponds to the presence of an additional coherent field mass, $M_{coh}$, proportional to $N'^2$. It is larger than the mechanical kinetic (ordinary) mass m by the factor $N'r_s$.* This factor may often be much less than unity, as we show below even for a system as large as a galaxy. However, for very large cosmic streams, for huge material flows of galaxies on the scale of the observable universe *or beyond*, this factor may become noticeably larger than unity, as we see below. *The additional coherent field mass $M_{coh}$ is in turn, also the source of a gravitational field in GR, the mass of which adds again to the total gravitational field mass, iteratively, thus defining a total coherent field mass $M_{dark}$. This important "Additional Nonlinearity," discussed in the next section, contributes to an effective enhancement of the factor $N'r_s$, allowing for the definition of a $N'r_{sEff}$ and of an $M_{dark} > M_{coh}$.* The incoherent gravitational field contributes to the mass of the particles (e.g., planetary systems including the star), and its ordered-incoherent (negative) energy contribution, proportional to N′, is contained in the mechanical kinetic energy of the ordered motion of the particles. It is present in the denominator of the first form of Eq. (2), and also in the energy of the disordered motion (e.g., of stars).

We consider a simple Lorentz transformation along the streaming direction from the local frame of the system of particles with low drift velocity v, to a frame at rest with the surroundings. This transformation was expected to only yield the drift mechanical kinetic energy of the particles, N′mv$^2$/2 to lowest order in v. However, from Eq. (2), we expect the same Lorentz transformation, *to also lead to a N′$r_s$ times "larger" coherent kinetic field energy* (also quadratic in v in lowest order. Usually N′$r_s$ << 1 at lower scales than the observable universe, as shown by Eq. (10) below.) By comparing the coefficients of v$^2$, we notice the presence of an additional mass of coherent gravitational nature, arising as a collective field mass, in fact from GR (see Sect. 2.3, 2.5, 2.6, 3.1–3.5). We may call this the coherent mass, the key for understanding the DM. This is in fact the gravitational (metric curvature) field energy of the large system of particles or objects, as we explain below. It must be a coherent mass, present also in the rest system, coherent at all scales, as we reason. It should be a result of Einstein's field equations, Eq. (6), as we show in Sect. 2.3–3.4. We may therefore drop the qualification "kinetic" for this mass.

We do not call the coherent gravitational coupling *different,* or *"non-minimal,"* to distinguish it from the coupling through the **g** field in the rest frame. They are the same GR coupling, seen from different frames of reference, related through Lorentz transformation. (see Appendix 1).

The result s″ = N′$r_s$ for the coherent gravitational field energy, or DM fraction, was derived in Eq. (2) in cylindrical flow geometry for a very long straight column of particles. Since the universe is considered flat, (see after Eq. 6) with the observable universe embedded homogeneously, our derivation would be applicable for the observable universe. Indeed, the observable universe would then be seen as part of a very long column of particles, moving in axial direction with a virtual, arbitrarily low speed v, which drops out from Eq. (2).

Considering on the intermediate scale the observed reticular cosmic web structure of the observable universe, a small region in any of the long branches of dark matter decorated with galaxies and other forms of baryonic matter, could also be considered to have cylindrical symmetry, for the purpose of approximating its coherent gravitational field energy.

On all lower scales, N′$r_s$ will over-estimate the DM fraction at long distances from the center of the structure, and under-estimate it at very small distances. This follows from comparing the 1/r dependence with a 1/r$^2$ decrease at distance r. We shall still use N′$r_s$ even at lower scales as a first approximation for estimating upper limits for the field energy, and eventually replace it with the much better calculation that uses our "key" $\rho = \lambda gg′$ for the DM density $\rho$ in Part II below.

### 2.2 Dark Matter Fraction in the Observable Universe

Consider the observable homogeneous, isotropic universe with size $R_u = 46.5 \cdot 10^9$ light years [4] in all directions, i.e., $4.4 \cdot 10^{28}$ cm, containing about $N = 2 \cdot 10^{12}$ galaxies [5, 6]. In cylindrical flow symmetry, an approximation of the number of galaxies per unit length, N′, in an equatorial plane of the observable universe is obtained by multiplying the number density in galaxies/cm³ with the area $\pi R_u^2$ of that equatorial plane

$$\begin{aligned} N' &= [2 \cdot 10^{12}/1.33\pi R_u^3][\pi R_u^2] \\ &= 6 \cdot 10^{12}/4R_u = 1.5 \cdot 10^{12}/4.4 \cdot 10^{28} \\ &= 3.4 \cdot 10^{-17} \text{ cm}^{-1}. \end{aligned} \quad (3)$$

Considering a representative mass of a galaxy [7, 8] to be $M = 10^{45}$ g, like our Milky Way, we obtain

$$r_{s\,gal} = 2GM/c^2 = 13.3 \cdot 10^{-8} \cdot 10^{45}/9 \cdot 10^{20} = 1.48 \cdot 10^{17} \text{cm}. \quad (4)$$

Thus,

$$N' r_s = 3.4 \cdot 10^{-17} \text{ cm}^{-1} \cdot 1.48 \cdot 10^{17} \text{ cm} = 5.032 \approx 5. \quad (5)$$

This result shows that the expected DM in the observable universe could be about five time the amount of ordinary matter noticed, or visible, in various forms: galaxies, including stars planets, gas and clouds of dust. *This is in agreement with the cosmic observations* [1]. ***In fact, we have defined here DM as the additional ordinary matter that would be needed to explain classically the observed gravitational fields.***

#### 2.2.1 Additional Nonlinearity

At first sight, this evaluation and identification of the large scale coherent field energy of the universe as the mysterious DM seems hard to reconcile with the known inhomogeneous distribution of DM along the branches of the observed cosmic web, trapping galaxies, gas and dust that decorate the web. However, the *additional iterative nonlinearity* mentioned in the preceding section, will cause a large field enhancement in places where visible matter and fluctuations have initially created a stronger field. This nonlinearity may yield agreement with the number 5 even for a lower average galaxy mass in Eq. (4). The uniform distribution of DM that could have perhaps been expected initially from our calculation, would have been unstable against small perturbations, due to the attractive nature of gravitation that acts also on the field itself according to Einstein's equations (EE), i.e., due to the gravitational field generating its own gravitational field. The latter contributes to the curvature of 4-dimensional space, expressed by the Riemann-Christoffel curvature tensor $R^\kappa_{\lambda\mu\nu}$, by the Ricci curvature tensor $R^\kappa_{\kappa\mu\nu} = R_{\mu\nu}$, and by *the curvature scalar* $R^\mu_{\ \mu} = \mathscr{R}$. *The latter includes a contribution from the local density and motion of*

*DM, which appears in the total energy-momentum tensor $T_{\mu\nu}$. Repeated indices are summed over.*

Einstein's field equations of general relativity can be written as

$$R_{\mu\nu} - g_{\mu\nu}\mathscr{R}/2 + \Lambda g_{\mu\nu} = 8\pi G T_{\mu\nu}/c^4, \tag{6}$$

where $T_{\mu\nu}$ is the energy–momentum density tensor of matter and field of any nature. $\Lambda$ is the cosmological constant, describing the pressure-like effect of "dark energy." The latter is needed to explain the observed acceleration of the Hubble expansion of the universe. As a property of vacuum, dark energy flattens the universe, compensating over the largest scale the curvature of spacetime introduced by ordinary and dark matter. The latter is calculated by us here as the coherent gravitational field of the universe, i.e. as a curvature of space–time. The gravitational field potentials, in form of the metric coefficients $g_{\mu\nu}$, are actually present on both sides of Eq. (6). *This causes the peculiar self-interaction of the gravitational field, a self-amplification at largest scales, where the field becomes dominant, also allowing for the accretion of coherent gravitational field mass (generated by the universe) by the weaker field of smaller structures, causing also field clumping and collapse processes.* Making from now on abstraction of $\Lambda$, which describes the "dark energy," and contracting with $g^{\mu\nu}$, we obtain

$$-\mathscr{R} = 8\pi G \mathscr{T}/c^4, \tag{7}$$

where $\mathscr{T} = T^{\mu}_{\ \mu} = g^{\mu\nu}T_{\mu\nu}$.

The numerator in Eq. (2) represents the coherent kinetic gravitational field energy, or DM kinetic energy, generated by the ordinary matter in motion. It represents a part of the $t_{00}$ component, of the $t_{\mu\nu}$ "tensor" density of the gravitational field, which will be defined and evaluated in Sects. 2.3, 2.5, and 2.6. This will show, at least in a Lorentz covariant way, how the density of DM ties in. The quotation marks express this limited covariance, restricted to quasi Minkowskian coordinates, that approach the Minkowski metric far from the system considered.

*The coincidence with the number 5 can be considered to be fortuitous, since the Milky way is above average in mass, but the order of magnitude obtained is our most interesting result.* Indeed, we can start again in cylindrical flow symmetry from the known critical density of matter in the universe, $\rho_{crit} = 9.47 \cdot 10^{-30}$ g/cm³, of which about 5%, i.e., about $\rho_o = 4.74 \cdot 10^{-31}$ g/cm³ is found to be ordinary matter. Then, $M_u = \rho_o 4\pi R_u^3/3$, (with $R_u$ as the radius of the observable universe), approximates its mass. Dividing by N, the estimated total number of galaxies, we obtain the average mass of a galaxy $M_g = \rho_o 4\pi R_u^3/3N$, and its Schwarzschild radius $r_s = 8G\rho_o \pi R_u^3/3Nc^2$.

Following the approximation used in Eq. (3) and dividing again $N$ by $4R_u/3$ to get $N'$, we obtain $N'r_s = 2G\rho_o \pi R_u^2/c^2$. With $R_u = 45 \cdot 10^9 ly = 4 \cdot 10^{28}$ cm, we finally get

$$\begin{aligned} s'' &= N'r_s \\ &= 1.33 \cdot 10^{-7} \cdot 4.74 \cdot 10^{-31} \cdot \pi \cdot 16 \cdot 10^{56}/9 \cdot 10^{20} = 0.35. \end{aligned} \tag{8}$$

*We notice that s'' is in fact not dependent on both the number and size of galaxies, and can be related directly to the contribution of ordinary matter to the critical*

mass density of the universe. Our coherence parameter $N'r_s$ can actually be written as $s'' = \Sigma_i N'_i r_{si}$, where the index i denotes the number $N'_i$ per length unit, of various groups of galaxies with differing masses and Schwarzschild radius values $r_{si}$.

The very largest scale, of the order of the observable universe, is essential for our result in Eq. (5). *Indeed, if we consider only a single galaxy, e.g., the milky way, we show here that the additional coherent gravitational field mass generated locally by the galaxy is negligible compared with the mass of the galaxy. The actual amount of DM present in the galaxy will thus include mainly DM generated by the universe, and attracted by the galaxy in varying amounts, like a foreign substance.*

As an example, let us thus evaluate in cylindrical flow symmetry the fraction of the own coherent gravitational field energy or DM, generated locally by the ordinary matter, mostly visible, in a galaxy alone. With about $10^{11}$ stars moving along a $10^4$ light years thickness of the disk-shaped galaxy, in a flow along the axis of the galaxy, we obtain

$$N' = 10^{11}/10^4 \text{ly} = 10^{11}/[10^4 \text{ly} \cdot 9.5 \cdot 10^{17} \text{ cm/ly}]$$
$$= 10^{11}/9.5 \cdot 10^{21} \text{cm} = 1.05 \cdot 10^{-11} \text{ cm}^{-1}. \quad (9)$$

Considering the mass of the sun $M = 2 \cdot 10^{33}$ g, as representative for the mass of a star, we obtain the Schwarzschild radius of a star to be

$$r_s = 2GM/c^2 = 2 \cdot 6.67 \cdot 10^{-8} \cdot 2 \cdot 10^{33}/9 \cdot 10^{20} \text{ cm}$$
$$= 3 \cdot 10^5 \text{cm} = 3 \text{ km} \quad (9a)$$

This yields a ratio of coherent gravitational field mass, created in the galaxy, to ordinary mass, given by

$$N'r_s = 10^{-11} \text{ cm}^{-1} \cdot 3.10^5 \text{ cm} = 3 \cdot 10^{-6}. \quad (10)$$

When the flow is considered realisically along the $2 \cdot 10^5$ ly diameter of the galaxy (or for a smaller galaxy), the result obtained will be even 20 times smaller. This estimation is qualified by the observation at the end of the preceding Sect. 2.1, on the limited validity and likely overestimation connected with the parameter $N'r_s$.

*This indicates that the coherent gravitational field mass created specifically by a single galactic flow of ordinary matter is indeed negligible.* However, we expect also a much larger coherent gravitational mass generated by the global flow, represented by the observable universe, and attracted in variable proportions into the respective galaxy, in a way "*guided*" by the small amount generated in the galaxy**.** *According to Newton's law of universal gravitation, the force on universe-generated DM of density proportional to $g'^2$, which can be attracted into a galaxy, is proportional with both: the universe's DM density~$g'^2$, and the field of the galaxy g. Indeed, the field g is defined as the force acting on the mass unit. In the local frame Newtonian physics applies. We will assume that the advected DM density is proportional to this force density and can be approximated by cubic terms like $gg'^2$.* This "guidance" or "amplification" of the small gravitational field "*g,"* generated by a single galaxy, will be caused through the nonlinearity of EE, in which the field interacts with itself, and also generates its own field. It can be understood like the coherent addition of a large term *g'* to a small term *g*

in a higher order term of a series expansion, e.g., a cubic term $(g+g')^3 = g'^3 + 3gg'^2 + 3g^2g' + g^3$. This results in terms like *$3gg'^2$* that amplify the energy density caused by the original small term g, contributing a much larger term $3gg'^2$, advected into the galaxy, *proportional to g; in general this is an advected DM concentration*

$$\rho = \lambda g g'^2, \tag{11}$$

*with a proportionality coefficient λ*. This creates the appearance, as if only the source of *g* (i.e., the baryonic matter in a galaxy) would be present, creating and commanding the whole dark matter amount present in a galaxy. *All DM is assumed to be coherent with itself, at most levels, as the coherent gravitational field of the observable universe*. As we shall see below, this explains the peculiarities seen in the well-known Tully–Fisher and Faber-Jackson relations, as well as the rotation curves of stars in galaxies: the baryonic make-up of the galaxy determines the larger DM effects. Furthermore, being proportional to the small gravitational field **g** generated by the galaxy itself, the total DM concentration in the galaxy *will not show a peak in the centrum of the galaxy where g is zero. DM avoids the center of mass*. A DM peak, or accumulation, was predicted by ΛCDM models and simulations in the center of galaxies. This absence of a DM peak is consistent with the observed Newtonian rotation velocities of stars in this central region. Nonlinearities higher than quadratic, e.g., cubic, known to be present in EE, will shape this amplification of the own gravitational field of the galaxy. While *g* represents the galaxy's or cluster's own gravitational field in this estimation, and $g^2$ is a measure of its own generated DM density, the term $g'^2$ corresponds to the much larger, more uniform DM background density of the universe, practically constant in the galaxy and in its DM halo region. The large amount of universal DM attracted into the galaxy could thus be considered to be proportional to $gg'^2$.

The result in Eq. (5) suggests that DM could exceed the visible matter in galaxies. This order of magnitude calculation shows how the observed fluctuating ratio of dark to visible matter in the universe can be also understood at the local level. Practically all of the DM found in galaxies is not locally generated, but is attracted into the galaxy, as we mentioned, being generated at the level of the observable universe. The fact that DM can be physically independent and separated from ordinary matter like a mysterious substance avoiding the center of mass, is noticed clearly [9] in the Bullet cluster. Indeed, In the Bullet cluster about 2/3 of the mass is in the hot gas that emits X-rays. That is where the center of mass of this structure is, after the collision with the smaller "bullet" part happened about 150 million years ago. The dark matter created at the level of the universe is attracted into the cluster proportional everywhere, to the much smaller own gravitational field of baryonic matter in the cluster. This field avoids the center of mass, and is present [9] on the periphery and in front of the bullet part, in the direction of its motion, where the total own gravitational field of both, the hot gas and the stars that just went through in the collision, is maximal. It is also present, with accreted DM, in a general peripheral halo. A more detailed calculation is possible with the "key" $\lambda gg'^2$ introduced in Eqs. (11) and (59) below, by first mapping the own gravitational field g of the baryonic matter in the cluster.

In a second step, using this mapping, the DM is calculated with density $\rho = \lambda g g'^2$ in every point. This would yield a DM concentration distribution that is very low in the vicinity of the center of mass, being close to zero there, while the baryonic concentration is finite, as can also be shown in individual galaxies. As shown for a single galaxy in Sects. 3.2–3.4 below, the DM will be in general present in a halo around the cluster system, where we also find the stars that passed the colliding hot gas region. The DM is observed indirectly also through its lensing effects.

### 2.3 General Relativistic Calculation in Cylindrical Symmetry

The ad-hoc expression, based on an electromagnetic analogy, proposed in the numerator of Eq. (2) for the coherent kinetic gravitational field energy may seem over-simplified. However, by writing Einstein's field equations in a form that is not manifestly covariant, we can express the energy–momentum "tensor" of the gravitational field of a matter flow system and its 00 component in a form that displays its dependence on $N'^2$ and $v^2$ that was obtained in Eq. (2). Indeed, consider a quasi-Minkowskian coordinate system, with a metric tensor $g_{\mu\nu}$ that reverts to the Euclidian-Minkowskian $\eta_{\mu\nu} = (1, -1, -1, -1)\delta_{\mu\nu}$. at infinity, at large distances. This metric can be written in the form

$$g_{\mu\nu} = \eta_{\mu\nu} + h_{\mu\nu}, \tag{12}$$

where $h_{\mu\nu}$ vanishes far from the material system considered, but is not considered very small. *This allows to define an energy–momentum "tensor" density $t_{\mu\kappa}$ of the gravitational field in terms of the departures $h_{\mu\nu}$ from the Minkowski metric.* From the definition of the curvature tensor $R_{\lambda\mu\,\nu\kappa}$ and of the Ricci tensor $R_{\mu\kappa} = g^{\lambda\nu}R_{\lambda\mu\nu\kappa}$ with Eq. (12) we can derive series expansions with respect to $h_{\mu\nu}$. Therefore, as shown in Eq. 7.6.14 of Weinberg's well-known book [10], the energy–momentum tensor of the gravitational field can also be expanded in terms of $h_{\mu\nu}$ and starts with quadratic terms in $h_{\mu\nu}$:

$$t_{\mu\kappa} = (1/8\pi G)\left\{-(1/2)h_{\mu\kappa}R^{(1)\lambda}_\lambda + (1/2)\eta_{\mu\kappa}h^{\rho\sigma}R^{(1)}_{\rho\sigma} + R^{(2)}_{\mu\kappa} - (1/2)\eta_{\mu\kappa}\eta^{\rho\sigma}R^{(2)}_{\rho\sigma}\right\} + O(h^3). \tag{13}$$

Here $R^{(1)}_{\rho\sigma}$ is the part of the (Ricci) curvature tensor linear in $h_{\mu\nu}$,

$$R^{(1)}_{\mu\kappa} = (1/2)\left[\partial^2 h^\lambda_\lambda/\partial x^\kappa \partial x^\mu - \partial^2 h^\lambda_\mu/\partial x^\lambda \partial x^\kappa - \partial^2 h^\lambda_\kappa/\partial x^\mu \partial x^\lambda + \partial^2 h_{\mu\kappa}/\partial x^\lambda \partial x_\lambda\right]. \tag{14}$$

$R^{(2)\rho\sigma}$ is the second-order part, and repeating indices are always summed over. In Eq. (13), the second-order part of $R^{\mu\kappa}$ is given by the subsequent Eq. 7.6.15 in the same book [10].

$$R^{(2)}_{\mu\kappa} = (1/2)h^{\lambda\nu}\left[\partial^2 h_{\lambda\nu}/\partial x^\kappa \partial x^\mu - \partial^2 h_{\mu\nu}/\partial x^\kappa \partial x^\lambda - \partial^2 h_{\lambda\kappa}/\partial x^\nu \partial x^\mu + \partial^2 h_{\mu\kappa}/\partial x^\nu \partial x^\lambda\right]$$
$$+ (1/4)\left[2\partial h^\nu_\sigma/\partial x^\nu - \partial h^\nu_\nu/\partial x^\sigma\right]\left[\partial h^\sigma_\mu/\partial x^\kappa + \partial h^\sigma_\kappa/\partial x^\mu - \partial h_{\mu\kappa}/\partial x_\sigma\right]$$
$$- (1/4)\left[\partial h_{\sigma\kappa}/\partial x^\lambda + \partial h_{\sigma\lambda}/\partial x^\kappa - \partial h_{\lambda\kappa}/\partial x^\sigma\right]\left[\partial h^\sigma_\mu/\partial x_\lambda + \partial h^{\sigma\lambda}/\partial x^\mu - \partial h^\lambda_\mu/\partial x_\sigma\right]. \quad (15)$$

In our electromagnetic analogy, the terms of $t_{00}$ in curly brackets in Eq. (13) are the gravitational field counterparts of the electromagnetic energy density

$$(1/2)[\rho V + (\mathbf{j}/c)\mathbf{A}] = (-1/8\pi)[V\Box^2 V + \mathbf{A}\Box^2\mathbf{A}]. \quad (16)$$

For the cylindrical flow system considered by us, the second (kinetic) term in these analogue expressions (16) is expected to be quadratic in **v** and in the particle concentration n′=N′/S, where S is the cross section of the flow and N′ again the number of similar objects per unit length in the direction of flow. A similar set of terms proportional to $n'^2$ and $v^2$ are expected to be present in Eq. (13). From symmetry we do not expect terms linear in v for $t_{00}$ in Eq. (13).

One may start from the static Schwarzschild solution for the metric of a mass m, with isotropic coordinates and with units yielding G=c=1 numerically

$$ds^2 = \left[(1-A_0)/(1+A_0)\right]^2 dt'^2$$
$$- (1+A_0)^4\left(dx'^2 + dy'^2 + dz'^2\right), \text{ with} \quad (17)$$
$$A_0 = m/2r_{\text{rest}}.$$

If seen from a coordinate system

$$t = (t' + vx')(1-v^2)^{-1/2} \quad y = y' \quad (18)$$

$$x = (x' + vt')(1-v^2)^{-1/2}; \quad z = z' \quad (19)$$

moving with the velocity v, with $\gamma = (1-v^2)^{-1/2}$, this becomes

$$ds^2 = (1+A)^2(dt^2 - dx^2 - dy^2 - dz^2) - \left\{(1+A)^4 - (1-A)^2/(1+A)^2\right\}\gamma^2(dt - vdx)^2. \quad (20)$$

Here we denoted $A|_{v=0} = A_0$ in Eq. (17) and

$$A = \left[mn'(1-v^2)^{1/2}/2\right]\left\{(x-X-vt)^2 + (1-v^2)\left[(y-Y)^2 + (z-Z)^2\right]\right\}^{1/2}$$
$$= mn'A'(v^2, x-X-vt, y-Y, z-Z); \quad A' = A/mn' \quad (21)$$

We have switched here from m to the mass density mn′ and we are considering now a three-dimensional flow of particles of mass m and concentration n′. Here the position of mn′ at t=0 is at X, Y, Z. Equations (20) and (21) give the metric coefficients generated at x,y,z by the mass per unit volume. From Eq. (20) one identifies the metric deviations "tensor" $h'_{\mu\nu}$:

$$h'_{00} = A^2 + 2A - [(1+A)^4 - (1-A)^2/(1+A)^2]\gamma^2; \tag{22}$$

$$h'_{11} = -A^2 - 2A - [(1+A)^4 - (1-A)^2/(1+A)^2]v^2\gamma^2; \tag{22a}$$

$$h'_{22} = -A^2 + 2A$$

$$h'_{33} = -A^2 + 2A$$

$$h'_{01} = h'_{10} = v[(1+A)^4 - (1-A)^2/(1+A)^2]\gamma^2 = v_1 F(a); \tag{22b}$$

$$\begin{aligned} h'_{02} &= h'_{20} = 0 \\ h'_{03} &= h'_{30} = 0 \\ h'_{12} &= h'_{21} = 0 \\ h'_{32} &= h'_{23} = 0 \\ h'_{31} &= h'_{13} = 0. \end{aligned}$$

Here $v_1 = v$ is the x-component of v. In a way similar to the derivation of the Coulomb potential and field at a distance r from an infinite line of charge, we define in the weak field limit the metric deviations $h_{\mu\nu}$ for a column of n′ particles of mass m per unit volume, and N′ per unit length, in cylindrical symmetry, i.e., for the flow we considered

$$\begin{aligned} h_{\mu\nu} &= \int dh'_{\mu\nu} = \int (\partial h'_{\mu\nu}/\partial M) dM \\ &= m \int (\partial h'_{\mu\nu}/\partial A)(\partial A/\partial m) n' dX dY dZ \\ &= mN' h''_{\mu\nu}(A', v, v^2). \end{aligned} \tag{23}$$

We used here the mass density $dM/dXdYdZ = mn'$, which yields the mass per unit length in the X direction by integration over Y and Z. We notice from Eq. (21) that the first power of v occurs in A through the expression x-X-vt. However, we integrate over X from minus infinity to plus infinity along the column, and over Y and Z over the cross section S, in Eq. (23). By integrating over X to get $h_{\mu\nu} = mN'h''_{\mu\nu}$, we lost the dependence on the first power of v. Therefore, with the exception of $h_{01}$ and $h_{10}$, these cylindrically symmetric integrated $h_{\mu\nu}$ will only depend on $v^2$ and no longer explicitly on v. Also, we see from Eqs. (21)-(23) that h″ or $A' = \partial A/\partial(mn')$ do not depend considerably on m or N′ in Eq. (23). They will remain small when m and N′ are scaled up to the size of the universe.

Equations (23) for $h_{\mu\nu}$ are now substituted into Eqs. (14), (15) and for $t_{00}$ in Eq. (13). Expanding the energy density $t_{00}$ in a series in terms of v for v << 1, we omit the first power based on time symmetry:

$$t_{oo} = t_{00}(0) + (v^2/2)(\partial^2 t_{00} \partial v^2)\big|_{v=0} + \text{higher powers of v.} \quad (24)$$

From Eqs. (15), (14) and (13) we see that, to lowest order in $h_{\mu\nu}$, the energy–momentum tensor $t_{\mu\nu}$ is bilinear in $h_{\mu\nu}$ and therefore, proportional to $N'^2$. By integrating over the whole space in and around the matter flow and by introducing a notation with T″,

$$(v^2/2)\int(\partial^2 t_{00}\partial v^2)\big|_{v=0} 2\pi r dr = 2G^2 m^2 (v/c)^2 N'^2 T'', \quad (25)$$

we notice that T″ is independent of m, N′ and v. It remains small if the flow is scaled up to the size of the universe. We can thus write the ratio between the coherent kinetic field gravitational energy and the mechanical kinetic energy, with units restored, to second order in v, in the form

$$s'' = s_g = 2Gm^2v^2N'^2T''/c^2[N'mv^2/2] = 2N'GmT''/c^2 = N'r_sT''. \quad (26)$$

This reproduces Eq. (2) with the corrective factor T'' that remains of the order of unity, even including logarithmic terms, when we scale the flow up to the size of the universe. This verifies our result, expressed in Eq. (2) at the level of the energy density, and concludes our general relativistic consideration for cylindric symmetry. We obtain this way our results in Eqs. (2)–(11) both ways. In Sects. 2.5 and 2.6 below we obtain a similar result in static spherical symmetry in the co-moving frame. This is done classically in Sect. 2.5 and in lowest order GR in Sect. 2.6.

## 2.4 Discussion of the Ansatz (v × g)²

We mention that in the last term of $R_{00}^{(2)}$ in Eq. (15), thus also in $t_{00}$ of Eq. (13), for a column of similar objects in stationary laminar flow along an arbitrary axis, not necessarily along the x-axis considered so far, we find terms corresponding to our Ansatz **v** × **g**. These terms are most important for the increase of $t_{00}$ above v = 0, at low v,

$$(-1/4)[\partial h_{\sigma 0}/\partial x^\lambda - \partial h_{\lambda 0}/\partial x^\sigma][\partial h_0^\sigma/\partial x_\lambda - \partial h_0^\lambda/\partial x_\sigma]. \quad (27)$$

In stationary conditions we approximate $\partial/\partial x^0 = 0$, so σ and λ can't be zero in this summation. The remaining terms are like the square of a curl of the equivalent gravitational "vector potential" ($h_{10}, h_{20}, h_{30}$).

Indeed, the components $h_{01}$, $h_{02}$ and $h_{03}$ of **h** are in general proportional to the components of **v** as, e.g., in Eq. (22b), i.e., **h** = **v**F. We see that F = F(A) is a scalar function of A with F(0) = 0, with derivative F′(0) = 6 and F(A) ≈ 6A ≈ 3 m/$r_{rest}$. This is for large r ≈ $r_{rest}$, corresponding to small A, according to Eqs. (17) and (21), for spherical symmetry at very low v, with γ ≈ 1. The spherically symmetric gravitational field would then be approximated by **g** = -Gm**r₀**/r², with **r₀** the unit vector of **r**. For cylindrical symmetry, we expect to get $F(A) = 3mN' \ln(r_{rest}) + const$. This is proportional to the cylindrical gravitational potential, which has the gradient -**g**. The curl of **h**($h_{01}$, $h_{02}$, $h_{03}$) is thus curl**h** = - **v** × grad F = -3mN' **v** × **r**$_o$/r. This is obviously proportional to **v** × **g**., as in Eq. (2).

More detailed derivations of the magneto-gravitational effects, used in our Ansatz of Eq. (2) can be found in the literature [11]. Equation (2) was first introduced by Handel in 2003 [3].

### 2.5 Classical Derivation of the Dark Matter Fraction of the Universe in Spherical Geometry

So far we have considered the observable universe as a column in flow through the rest of the universe, and found the coherent gravitational field fraction to be $N'r_s = 0.75 r_s N/R_u$ in Eqs. (3)–(5). In a way similar to the static electromagnetic analogy shown in Appendix 1, we consider here a static observable universe, with spherical symmetry, first classically, then in GR. We calculate the fraction of coherent gravitational field energy/mass to the energy/mass known to be present in the ordinary (mainly baryonic) matter of the observable universe. This calculation is done in the rest frame of the observable universe.

We consider for simplicity again $N = (4\pi a^3/3)n$ galaxies, all of mass m, to be present with a uniform density n in the observable universe of radius $R_u = a$. From the Gauss theorem, the gravitational field is

$$\mathbf{g} = -(4\pi/3)Gnm\mathbf{r}; \quad r \leq a \tag{28}$$

$$\mathbf{g} = -(4\pi/3)Gnm(a^3/r^2)\mathbf{r}/r \quad r > a \tag{29}$$

The classical energy of the gravitational field is

$$\begin{aligned} W_{int} &= (1/8\pi G) \int_0^a \mathbf{g}^2 4\pi r^2 dr \\ &= 2(\pi/3)^2 G n^2 m^2 \int_0^a r^4 dr = 0.1 \, GN^2 m^2/a \end{aligned} \tag{30}$$

$$\begin{aligned} W_{ext} &= (1/8\pi G) \int_a^\infty \mathbf{g}^2 4\pi r^2 dr \\ &= 2(2\pi/3)^2 G a^6 n^2 m^2 \int_a^\infty dr/r^2 = 0.5 \, GN^2 m^2/a \end{aligned} \tag{31}$$

The total rest energy of the field is thus,

$$W_{total} = 0.6 \, GN^2 m^2/a \tag{32}$$

This can be compared with the rest energy of the ordinary matter, $Nmc^2$. Dividing by $Nmc^2$, we obtain this way, similar to Eq. (2),

$$s'' = (0.1 + 0.5) \, GN^2m^2/amNc^2 \\ = 0.8(3N/4R_u)r_s = 0.4N'r_s. \tag{33}$$

Here we defined N′ the same way as in Eqs. (2)–(5), as $N' = 3N/4R_u$. We recover tis way the results obtained in cylindric flow symmetry with a corrective factor of 0.4. We thus find the result $s'' = N'r_s = 5$, seen in Eq. (5), being replaced in our classic calculation in spherical symmetry, by $0.4·5 = 2$ i.e., $s'' = 2$ in Eq. (33). We also find the result $s'' = N'r_s = 0.35$, seen in Eq. (8), replaced here, in static spherical symmetry, by 0.14, i.e., $s'' = 0.14$. As we show in the next section, a similar result can be obtained by solving EE in spherical symmetry in lowest order. This similarity with Newtonian physics is used in Part II to describe the advection of DM generated at the level of the universe into lower-scale structures.

## 2.6 Relativistic Calculation of the Fraction of Dark Matter in the Universe in Spherical Geometry

We start again with the Schwarzschild solution in isotropic coordinates, Eq. (17) that yields

$$g_{00} = (1-A)^2/(1+A)^2; \; h_{00} = (1-A)^2/(1+A)^2 - 1, \\ g_{11} = g_{22} = g_{33} = -(1+A)^4; \\ h_{11} = h_{22} = h_{33} = 1 - (1+A)^4; \; A = mN/2r = r_s/4r \tag{34}$$

Other $g_{\mu\nu}$ and $h_{\mu\nu}$ are zero. All this is for $r \geq a = R_u$. For $r < R_u$, the interior Schwarzschild solution must be used.

Guided by the classical solution obtained in the previous section, we estimate the gravitational field energy by integrating its density only over the exterior parts of the universe, $r > R_u$.

For distances $r > r_s/4$, we have $|A| < 1$ and we can expand the metric deviation $h_{00}$

$$h_{00} = -4A + 8A^2 - 12A^3 + 10A^4 + \cdots \\ h_{11} = h_{22} = h_{33} = -4A - 6A^2 - 4A^3 - A^4. \tag{35}$$

*To lowest order in A, we notice that all nonzero metric deviations $h_{\mu\nu}$ equal -4A.* Then, since Eqs. (13) and (15) are bilinear in $h_{\mu\nu}$ when terms of higher order than 2 in $h_{\mu\nu}$ are neglected, we can write, consistently limiting ourselves to $h_{\mu\nu}$ calculated only to first order in A,

$$t_{\mu\nu} = m^2 N_<^2 t''_{\mu\nu i} \quad (r < R), \tag{36}$$

$$t_{\mu\nu} = m^2 N^2 t''_{\mu\nu e} \quad (r > R), \tag{37}$$

where $N_<$ is the number N limited to a radial coordinate below r. Here, again the $t''_{\mu\nu}$ remain small, when m and N are scaled up to the order of magnitude of the whole universe. Equations (36)–(37) are important, in particular, for the energy density $t_{00}$ of the gravitational field.

In order to outline, or display, e.g., a first estimation of $t_{\mu\nu e}$ in spherical static geometry up to second order in $h_{\mu\nu}$, from Eq. (13), we limited ourselves in *Appendix 2* at large distances r, only to the first, term, − 4A, in Eqs. (34) and (35). With units restored, this yields a lowest order estimate

$$t_{00e} = 2Gm^2N^2/8\pi r^4. \tag{38}$$

One identifies $t_{\mu\nu e}''$ this way for $\mu = \nu = 0$ as

$$t''_{00e} = 2G/8\pi r^4. \tag{39}$$

This energy density can then be integrated over the outside region $r > R_u = a$, similar to the classical case in Eq. (31):

$$W_{ext} = \int_a^\infty t_{00e} 4\pi r^2 dr = Gm^2N^2/a \tag{40}$$

If we assume the internal gravitational energy to contribute another 20%, like in Eq. (30) and (31), we obtain

$$W_{tot} = 1.2\, Gm^2N^2/a \tag{41}$$

By comparison with the rest energy of ordinary matter $Nmc^2$, we obtain with $N' = 3N/4a$,

$$s'' = 1.2(4/3)Gm^2N'N/Nmc^2 = 0.8N'r_s. \tag{42}$$

***This is twice the classical result in Eq.*** (33)***.*** The corrective factor 0.4 of Eq. (33) is replaced by 0.8. This would yield 0.8·5 = 4 in Eq. (5). The result $s'' = N'r_s = 0.35$, seen in Eq. (8), would be replaced here, in static spherical symmetry, by, $s'' = 0.28$. Including all higher-order nonlinearities present in an exact solution of Eq. (6), it may be possible to obtain the observed DM fraction of 5 with a more realistic galactic mass. We thus gain a rough understanding of the relation between classical and low-order GR approaches, which inspires the semiclassical derivations in Part II below. It is also remarkable that, when applied to the observable universe, the cylindric-flow and spherical geometries yield the same order of magnitude, with some over-estimation in cylindrical symmetry.

## 3 Derivation of Rotation Curves, of the Dark Matter Density Formula and of the Tully–Fisher Relation with a Newtonian Approximation Ansatz

### 3.1 Analysis of the Dark Matter Problem

Any exact solution of EE, including all length scales of the universe, should define DM, the coherent gravitational field in the observable universe, and its distribution in structures at all lower scales. It should explain the observed structures and their motion. It should also explain the absence of DM phenomena at the lowest scale of planetary systems, their increasing presence at larger, galactic, scales, and their dominance at the scale of the observable universe. This dominance is expected, since the field generates its own field, more so at large scales and larger masses, where the self-interaction of the field becomes more important. In particular, this suggests the importance of the mentioned self-interaction for the largest-scale gravitational fields symbolically denoted with g′, which also causes the cosmic web.

These self-interacting gravitational fields g′ and the corresponding DM generated with concentration proportional to $g'^2$ at the largest scale of the universe, will also interact coherently, in this exact GR solution, with the own gravitational fields g of lower-scale structures, like galaxies and their clusters. At the largest scale, the self-interaction of the field is expected to become more important than its mutual interaction with ordinary matter, mainly baryonic. We have seen this demonstrated, and the dominance of the gravitational field discussed, in Part I, Sects. 2.2–2.6 in spherical and cylindrical flow symmetries, classically and in lowest order GR. This dominance requires other methods describing the strong nonlinearity, as we developed here in Sects. 3.2–3.4. The same idea, of nonlinear self-interaction of the dominant gravitational field, has been considered qualitatively before [12] in terms of the well-known mutual interaction of gravitons, in a weak field approximation. Here we treat this nonlinearity with an elementary physical Ansatz, first introduced in Eq. (11), which yields universal quantitative results, in startling, natural agreement with observations, allowing the calculation of the observed DM at all lower scales of the observable universe … with a classical, Newtonian, approximation applied to DM!

In Part I, Eq. (11) we have approximated this self-interaction of the largest scale field g′ with the fields g of lower-scale structures, through the formula $\rho = \lambda gg'^2$. This is our Newtonian approximation of the GR advection or accretion of universal DM $\sim g'^2$ on the field g of lower-scale structures.

Indeed, since g is the gravitational force acting on the mass unit, and since $g'^2$ is a measure of the energy or mass density, of the large scale field g′, we consider $gg'^2$ a measure of the force density attracting universal DM into the galaxy or other structures considered. **In our Ansatz, the advected DM density is simply considered proportional with the force density that causes it, with an effective physical coefficient** λ. The latter could be somewhat smaller for the few structures found in the voids of the cosmic web. It is startling to see how the complexity of Einstein's equations *is simplified again at the largest* *scale,* because of the exclusive dominance of DM, the gravitational field, which becomes mainly its own source, for the observable universe as a whole. We are using in fact Einstein's equations in the local frame, which revert to Newton's g and g′ fields and to the simple gravitational force notion, to the force density proportional to gg′$^2$ in the classical approximation. This creates the appearance that no GR would be involved at all in Sects. 3.1–3.4. In fact we are considering, in the local co-moving frame, the large-scale GR metric field, the gravitational field, which is mainly its own source in Eq. (6). For this largest scale regime we apply a Newtonian field approximation for the coherent advection of DM with density ρ = λgg′$^2$, enhancing the gravitational field g of smaller scale structures, e.g., galaxies and their clusters.

The local frame is restricted in general to the differential neighborhood of the co-moving point in space–time. However, if we consider the coherent gravitational field g′ of the universe relatively constant at the very much lower scale of a galaxy or a cluster of such galaxies, our Newtonian approximation could still find a limited application, in spite of the larger-scale non-uniformities of the cosmic web. There are about 2·10$^{12}$ galaxies estimated to be present in the universe.

Mathematically. such a simplification singles out a particularly important type of "third-order-in-g" interactions implied by GR, by Eq. (6) in the field-dominated domain, but neglected in Eq. (13). Physically it assumes that the accretion of DM generated at the largest scale is proportional to the force exerted by the gravitational field **g** of lower-scale structures on the large-scale DM density, whose measure is g′$^2$.

Our identification of the main nonlinearity, describing the large-scale gravitational field in Einstein's equations, can explain and derive the rotation curves of stars in galaxies, as well as the Tully–Fisher relation, and in general the amounts of DM present at lower scales, as shown in the next three sections.

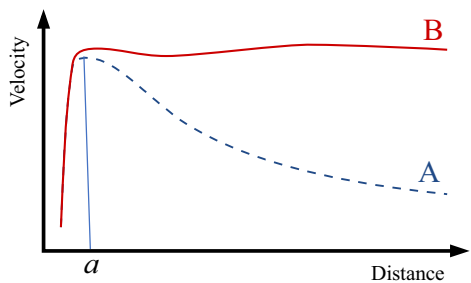

**Fig. 1** After reaching a maximum close to the Newtonian prediction, the same prediction suggested a decrease of the rotation speed of stars in galaxies at larger distances from the galactic center, as shown by curve A. The observed dependence is shown by curve B. By Phil Hibbs—his work in Inkscape 0.42, Public Domain, https://commons.wikimedia.org/w/index.php?curid=365013

### 3.2 Derivation of the Rotation Curves of Stars in Galaxies

The rotation speed of stars of mass m in the outer regions of spiral galaxies has been noticed to decrease slower than classically predicted with distance r from the galactic center. Instead of decreasing proportional to $1/r^{1/2}$ with the distance r from the galactic center (Curve A, in Fig. 1), after reaching the maximum predicted by Newtonian theory from the increase of the cumulative attractive mass with distance r in the central region, the rotation speed of stars remains practically constant, instead of decreasing with r; see the observed curve B in Fig. 1.

*Based on GR, our understanding of DM as the coherent gravitational field of the universe easily explains the rotation curve B and the nature, form, and properties of galactic halos.* Indeed, *let us assume* for simplicity that all the baryonic mass M of the galaxy is present at distances less than the first rotation speed maximum, which is at almost the same distance *a* on both curves in Fig. 1. According to Newtonian concepts, for r > *a*, the speed v of stars of mass m, rotating at distances r beyond the maximum in Fig. 1, should still be determined by

$$mv^2/r = GmM/r^2; \quad v^2 = GM/r. \tag{43}$$

Here M is the baryonic mass present in the galaxy. However as we mentioned before, at the end of Sect. 2.2, in general, DM generated by the universe with average concentration proportional to $g'^2$ is attracted into the galaxy that has its own field g, by gravitational attraction forces proportional to $gg'^2$. We can thus expect, based on the spirit of the GR field equations, the deposition of attracted DM from the universe to be also proportional with both the field $g = GM/r^2$ of the baryonic matter of the galaxy, and the universal DM concentration $g'^2$ that is constant over the whole region around the galaxy, varying over larger length scales. This calculation "key" corresponds to terms in $O(h_{\mu\nu}^3)$ in Eqs. (13), (75) and (83). We therefore expect a DM concentration given by the key $\rho = \lambda g'^2 g = \lambda g'^2 GM/r^2$ to be attracted and deposited proportional with the own gravitational field $g = GM/r^2$, (for r > a) and $= GMr/a^3$ (for r < a), that is generated by that galaxy. Here $\lambda$ is the coefficient of proportionality introduced by Eq. (11). We obtain thus, for r > *a*, a cumulative, spherically symmetric, DM and baryonic mass for the galaxy, $M_<$, present at distances less than r, given by

$$\begin{aligned}M_< &= M + Gg'^2 M\left[\int_0^a r^3/a^3 + \int_a^r\right](\lambda/r^2)4\pi r^2 dr \\ &= M + Gg'^2 M \int_0^a (\lambda r^3/a^3)4\pi dr \\ &\quad + 4\lambda\pi Gg'^2 M(r-a) \\ &= M - 3\lambda\pi Gg'^2 Ma + 4\lambda\pi Gg'^2 Mr.\end{aligned} \tag{44}$$

For r ≤ a we obtain, with a DM concentration $\rho = \lambda g'^2 GMr/a^3$,

$$\begin{aligned} M_< &= Mr^3/a^3 + 4G\lambda\pi g'^2 M \int_0^r r^3/a^3 dr \\ &= Mr^3/a^3 + G\lambda\pi g'^2 Mr^4/a^3. \end{aligned} \quad (44a)$$

The cumulative DM in Eq. (44) increases for $r > a$, proportional to r, while its concentration decreases like $1/r^2$. From Eq. (44) we obtain then for $r > a$, in the same approximation,

$$\begin{aligned} v^2 &= G(M_<)/r \\ &= GM[(1 + \lambda\pi Gg'^2 a)/r + 4\lambda\pi Gg'^2(1 - a/r)] \\ &= GM[(1 - 3\lambda\pi Gg'^2 a)/r + 4\lambda\pi Gg'^2]. \end{aligned} \quad (45)$$

For $r \leq a$,

$$v^2 = G(M_<)/r = GM\left[r^2/a^3 + G\lambda\pi g'^2 r^3/a^3\right]. \quad (45a)$$

By iteration this converges, because G is small. We notice that Eqs. (45) and (45a) reproduce the features of curve B in Fig. 1, provided $(d/dr)(v^2) \geq 0$ always. This happens when $3\lambda\pi Gg'^2 a \approx 1$, i.e., it is as large as (or just somewhat larger by a very small amount than) the unity. *We prove in Eq. (61) below, on the basis of the Tully–Fisher relation that this is indeed the case, with $3\lambda\pi Gg'^2 a = 1.083$.* The further slight buckling of curve B beyond $r = a$ may be caused by the presence of some stars even at $r > a$ in the arms of spiral galaxies, although we assumed for simplicity that all the baryonic mass M of the galaxy is present at distances less than the first rotation speed maximum, which is at almost the same distance *a* on both curves in Fig. 1: see Eqs. (47)–(52) for relaxing this assumption. We obtain this way a linear Newtonian increase of v for $r \leq a$ in Eq. (45a) (the DM part $\sim r^{3/2}$ of v increases slower, with vanishing value and vanishing slope at $r = 0$) *and an essentially constant speed at all larger distances.* This constant rotation speed could be observed up to a very large distance D, where the attraction from other galaxies reduces to zero the total gravitational field **g** of the galaxy (along with reducing to zero its remaining small DM halo density ρ). The total **g** reverses its sign, by entering the domain of other galaxies. Up to these large distances $D >> a$, (over $10^6$ ly for the Milky Way or Andromeda), the total accreted DM still increases, although its concentration has gradually decreased proportional to $1/r^2$ almost to zero. (Note: there may be a faster decrease of the halo density of the galaxy, right before D, if g gets below a certain very low background value $g_{oo}$).

By setting $r = D$ in Eq. (44), we obtain the total mass of the galaxy

$$M_t = M - 3\pi\lambda Gg'^2 Ma + 4\pi\lambda Gg'^2 MD, \quad (46)$$

and the total DM in the galaxy, approximated with $D >> a$,

$$M' = -3\pi\lambda Gg'^2 Ma + 4\pi\lambda Gg'^2 MD \approx 4\pi\lambda Gg'^2 MD. \quad (46a)$$

If we would set the background $g_{oo}$ to be zero, as a reasonable approximation, we would obtain from Eqs. (44a) and (46a) a roughly D/a times larger DM halo mass than the baryonic mass at r < a.

The calculation in Eqs. (43)–(45) considers the baryonic mass M of the galaxy confined with constant density *nm* in a central sphere of radius *a* that is negligible in size, compared to the distance D >> a to the domain of the next galaxy. This caused us to miss the slight bulge in curve B at r > a.

### 3.2.1 Improving the Assumed Galactic Mass Distribution

To improve on this approximation that considers all baryonic mass confined to r < a, we may consider instead a *more realistic* rounded-up continuous distribution of baryonic matter with a resulting field $g_o(r) = \alpha/[r^{-1} + (r^2/a^3)]$, with the same dependence on r as g, at both small and large distances,

$$g_o(r) = 4\pi G <nm> ra^3/3(r^3 + a^3) = \alpha r a^3/(r^3 + a^3), \tag{47}$$

where

$$\alpha = 4\pi G <nm> /3. \tag{47a}$$

Here <nm> is the average baryonic density $\rho_b$ for r < a. From the Gauss theorem (applied at r = ∞ and at r = a), we see that in this case, *a* is the median radius of the galactical baryonic mass distribution $\rho_b = -(1/4\pi G)\text{div}\mathbf{g_o}$: the same baryonic mass is located closer than *a,* and farther than *a*. Adding again universal DM with a density proportional to $g_o(r)$ like in Eq. (44), we obtain the total radial gravitational field strength g, generated in the galaxy

$$g = GM(r)/r^2 = \alpha r a^3/(r^3 + a^3) + (\alpha a \beta/r^2) \int_0^r 3a^2 r^3 dr/(r^3 + a^3). \tag{48}$$

where

$$\beta = 4\pi\lambda G g'^2/3. \tag{49}$$

The integral in Eq. (48) yields

$$J = 3a^2 r + a^3 \{\log(1 + r/a) - (1/2)\log[(r/a - 1/2)^2 + 3/4] - 3^{1/2}\arctan[3^{1/2}r/(r-a)]\}. \tag{50}$$

This yields for the rotation speed of baryonic matter in the galaxy at a distance r from the center

$$v^2 == \alpha r^2 a^3/(r^3 + a^3) + (\alpha a \beta/r) a^3 \{\log(1 + r/a) - (1/2)\log[(r/a - 1/2)^2 + 3/4] - 3^{1/2}\arctan[3^{1/2}r/(r-a)]\} + 3\alpha a^3 \beta. \tag{51}$$

The last, maximum speed term, in Eq. (51),

$$v_M^2 = (4\pi)^2 \lambda G^2 g'^2 nma^3/3 = 4\pi\lambda G^2 g'^2 M, \tag{52}$$

is present also in Eq. (45). With the right value of the parameter $\beta$ defined by Eq. (49), Eqs. (45) and (52) should reproduce the observed red curve in Fig. 1 even better.

The values of $\beta$ or $\lambda g'^2$ are determined by the observed coefficient present in the baryonic Tully–Fisher relation (53) in the next Sect. 3.3. As we also show in the next section, these values, determined this way, reproduce the observed rotation speed curve exactly, convincingly verifying our approach.

The often used Navarro–Frenk–White profile model [13] of the density of galactic DM halos as a function of r includes a term similar in form to our term $(\alpha\beta/r) a^4\log(1+r/a)$ implied in Eqs. (48)–(51). This model, however, is based on N-body simulations with hypothetical DM particles, which correspond to a different, assumed, physical situation.

### 3.3 Derivation of the Baryonic Tully–Fisher Relations and Details of the Rotation Curves; Universal Dark Matter Key

R. Brent Tully and J. Richard Fisher published an empirical relation between luminosity and the maximal rotation velocity of spiral galaxies in 1977. The relation is even better verified, when it is considered between the baryonic mass *M* of the galaxy (stars plus gas) and the velocity *v* of baryonic matter in outer regions [14] up to the much larger distance D, where the field reverses sign, pointing to a neighboring galaxy:

$$M/M_\odot = 58(v/1 \text{ km/s})^{3.97}. \tag{53}$$

Here $M_\odot$ is the mass of the sun.

As we have seen in Eqs. (46) and (46a), since D > > a, we expect the last (DM) term in Eqs. (44) and (46), representing matter in the halo, to be dominant for values of r approaching D. We can thus approximate the total galactic mass $M_<$ at distances r close to D in the Halo, with the last term in Eq. (44), i.e., with the DM halo mass. Setting r=D in Eq. (44), we obtain thus for the total mass of the galaxy including DM

$$M_t \approx 4\pi G\lambda g'^2 MD. \tag{54}$$

As we see from Eqs. (2), (5), (8), (33), (42), the DM fraction in the universe is about $s''$ times larger than the baryonic matter in mass, which is about 5 times, both experimentally and theoretically based on GR. For a single galaxy, $s''$ is the DM fraction generated as the own coherent gravitational field fraction, which is about $10^{-6}$ or less, as shown by Eq. (11). However, as mentioned before, DM generated at the level of the universe is attracted into the galaxy with local concentrations proportional to this gravitational field strength generated by the galaxy, yielding a much larger actual DM fraction. For spiral galaxies large enough to have a measurable rotational speed of (mainly hydrogen gas) matter in their DM halos, the Tully–Fisher Eq. (53) is observed. Our Milky Way is a good example of these galaxies, and will be considered representative, although the observations were done on other galaxies. For these galaxies we expect an actual total

DM fraction γ much larger than s″. We can estimate the DM fraction also from the empirical Tully–Fisher relation, which is derived here as follows.

The actual resulting DM fraction γ in a galaxy can be obtained like the DM fraction of the universe in Eq. (33), by dividing the actual resulting gravitational energy by the baryonic rest energy. We thus obtain for a spiral galaxy with baryonic mass M an approximation of the total gravitational field energy,

$$G(M_t)^2/D = \gamma Mc^2. \quad (55)$$

The actual (mainly accreted) DM fraction is given here by the parameter γ, similar to the s″ parameter in Eq. (33), applicable to the whole universe. The maximal rotational speed $v$ is then given by

$$v^2 = GM_t/D = [\gamma GMc^2/D]^{1/2}, \quad (56)$$

where a < r < D. In the last form we have substituted $M_t$ through its value obtained from Eq. (55). *This spells our derivation of the Tully–Fisher relation in the form*

$$\mathbf{v^4 = \gamma GMc^2/D}. \quad (57)$$

Comparing this with the observed Eq. (53), in which we approximate the exponent 3.97 with 4, we obtain in CGS units, e.g., by multiplying Eqs. (53) and (57), with D = $10^6$ ly = 9.5 $10^{23}$ cm,

$$\begin{aligned}\gamma &= D(1\text{ km/s})^4/58Gc^2M_\odot; \\ &= 10^6 \cdot 9.5 \cdot 10^{17} \cdot 10^{20}/58 \cdot 6.7 \cdot 10^{-8} \cdot 9 \cdot 10^{20} \cdot 2 \cdot 10^{33} \\ &= 1.36 \cdot 10^{-5}.\end{aligned} \quad (58)$$

This derivation of the Tully–Fisher relation and determination of the parameter γ allows for a calculation of the quantity $\lambda g'^2$, which controls the accretion of DM of the universe into the galaxy or any other structures at scales well below the whole universe. Indeed, by substituting Eq. (54) into Eq. (55), eliminating $M_t$, we obtain from Eq. (11), with $10^{11}$ average stellar masses of 0.07 solar masses each, in the baryonic mass M of a galaxy as in Eq. (8),

$$\begin{aligned}\lambda g'^2 &= (c/4\pi)(\gamma/MDG^3)^{1/2} = (3 \cdot 10^{10}/4\pi) \\ &\quad \times \left[1.36 \cdot 10^{-5}/10^{11}\ 1.4 \cdot 10^{32}\ 9.5 \cdot 10^{23}(6.7)^3\ 10^{-24}\right]^{1/2} \\ &= (2.39 \cdot 10^9)[3.4 \cdot 10^{-51}]^{1/2} = 1.39 \cdot 10^{-16}\ \text{gs}^2/\text{cm}^4.\end{aligned} \quad (59)$$

*Here λ is the DM accretion constant, which defines the universal key that allows us to calculate the amount of universal DM accreted by the own gravitational field of structures at lower scales below the observable universe in the spirit of GR.* Substituting this into the rotation curve Eq. (45) for r ≥ a, we obtain

$$v^2 = GM\left\{(1 - 3\pi\lambda g'^2 Ga)/r + 4\pi\lambda g'^2 G\right\}. \quad (60)$$

At r = a, the DM terms in curly brackets together yield a quarter of the last DM term in curly brackets. This is to be compared with 1/a, which comes from the first term in curly brackets, the baryonic contribution. With Eq. (59) and Milky Way parameters a = 1.3·10⁴ly (about the size of the central bar of the Milky way), we calculate

$$3\pi\lambda g'^2 Ga = 9.42 \cdot 1.39 \cdot 10^{-16} \; 6.7 \cdot 10^{-8} \; 1.3 \cdot 9.5 \cdot 10^{21} = 1.083. \quad (60a)$$

From Eqs. (60) and (60a) we obtain for r ≥ a:

$$v^2 = GM\{[1 - 1.083]/r + 1.444/a\}. \quad (61)$$

*Here indeed 1.083 > 1, i.e., $3\lambda\pi Gg'^2 a > 1$, only slightly larger, as suggested in* Sect. *3.2, right after Eqs.* (45) *and* (45a)*. It always yields a positive derivative of $v^2$, as required to reproduce the observed rotation curve in* Fig. 1*. Thus, the DM term in 1/r with λ in Eq.* (45) *is slightly larger than the term without λ. This can't just be a coincidence. It rather proves the fascinating precision and consistency of our semi-classical formula $\lambda gg'^2$.* **At r = a, DM is about 36% of the baryonic matter, both in cumulative mass and in contribution to $v^2/r$, in our $\lambda gg'$-based approximation of the exact solution of Einstein's equations.** Indeed, we also obtain from Eq. (45a) and (59) for r ≤ a:

$$v^2 = GM\{r^2/a^3 + \pi\lambda g'^2 G r^3/a^3\} = GM(r^2/a^3)[1 + 0.361 r/a]. \quad (62)$$

Equations (61 and (62) show the continuity at r = a, well-known from experiment, with a positive derivative $(d/dr)(v^2)$ everywhere, that goes to zero soon for r > a and has a discontinuity at r = a. **Equations** (61) **and** (62) **reproduce indeed the curve B in** Fig. 1. **It also proves the correctness of our DM theory and of the Ansatz $\lambda gg'^2$ that we have used to describe the advection of universal DM, as inferred from GR, into lower scale structures with own generated field g, e.g., into galaxies. The numerical consistency obtained is indeed very good.** We expect this to be borne out by later direct derivations from Einstein's nonlinear field equations. Note, however, that the slight buckling, visible in Fig. 1 at larger distances, can be caused by the presence of some stars in the region r > a. *On the other hand, the discontinuity of the derivative $d/dr(v^2)$ at r = a would be avoided by using a more realistic baryonic matter distribution like in Eqs.* (46)–(52).

In fact, the accretion of universal DM into the halos of galaxies should stop at very large distances, after the 1/r²—like decrease of the own gravitational field g, generated directly by the galaxy, drops below the small general background $g_{oo}$ that was mentioned before. The density of the accreted DM would then be $\rho = \lambda g'(g - g_{oo})$. A lowest limit of $g_{oo}$ is given by quantum fluctuations but the background could be much higher. This subtraction of $g_{oo}$ would not change our results, except in two cases:

(1) The evaluation of the total mass including DM mass of a galaxy, where the accretion would stop way before the field reversal caused by a neighboring galaxy. This limits the total accreted DM to about an order of magnitude above the galactic baryonic mass. In the Milky way this could be about 8–30 times. Indeed, from Eqs. (44a) and (46a) we would obtain about $M'/M = D/a = 10^6$ly $/1.3 \cdot 10^4$ly $= 77$ times more DM in the galaxy than baryonic matter, based on our approximation of GR.

(2) The evaluation of the DM amount in small structures, like the solar system (Sect. 3.4) and other planetary systems, where the obtained very small DM amount will be reduced even more.

In Eq. (8), the effective average Schwarzschild radius $r_s$ of a galaxy should thus be considered to be larger than our calculation assumed by including only baryonic matter, when the non-linearity of Eq. 6 is included correctly.

DM in the form of the gravitational field of the universe $g'$, creates more DM ($\approx \lambda g g'^2$), according to Einstein's nonlinear equations, where it is present both as a source and as a resulting field. This creates paradoxical situations at lower scales, in galaxies and in their clusters, where universal DM is accreted coherently in large amounts, according to GR, like mysterious halo substance. The nonlinearity causes the superposition principle, often still surreptitiously implied in our thinking, to fail.

### 3.4 Dark Matter in the Solar System

Knowing the mass of the sun $M = M_\odot = 2 \cdot 10^{33}$ g, its radius $a' = 7 \cdot 10^{10}$ cm and the distance to earth $r = 1\text{AU} = 1.5 \cdot 10^{13}$ cm $>> a'$, we can calculate with Eq. (11) the cumulative DM inside the earth's orbit, obtaining like in Eq. (44)

$$M_< = M_\odot - 3\lambda\pi G g'^2 M_\odot \cdot a' + 4\lambda\pi G g'^2 M_\odot \cdot r. \tag{63}$$

We use the "key" $\lambda g'^2 = 1.39 \cdot 10^{-16}$ gs$^2$/cm$^4$ from Eq. (59). Neglecting the middle term, we obtain in Eq. (63), at $r = 1$AU

$$M_< = M_\odot \cdot [1 + 4G\lambda\pi g'^2 \cdot 1\text{AU}] = M_\odot[1 + 4 \cdot 6.7 \cdot 10^{-8}\pi \cdot 1.39 \cdot 10^{-16} 1.5 \cdot 10^{13}]$$
$$= \{1 + 1.751 \cdot 10^{-9}\} M_\odot. \tag{64}$$

This shows that DM is present, but at the earth's orbit, it only corresponds to a negligible increase of the mass of the sun. For the increase of the quadratic rotation speed of the earth we obtain with $r = D' = 1$AU the same small fractional increase

$$v^2 = G(M_<)/D' = GM_\odot(1 + 4\lambda\pi G g'^2 D')/D' = (GM_\odot/D')(1 + 1.751 \cdot 10^{-9}). \tag{65}$$

For the speed of the earth on its orbit this fractional increase would amount to $0.9 \cdot 10^{-9}$. This may be too small to be verified experimentally, with the uncertainty in the mass of the sun. However, the corresponding small fractional increases in the orbital velocities of Jupiter and of the outer planets are proportional to their distances from the sun. This may cause very small phase differences over time, eventually observable in their positions on their orbits. Furthermore, the effects that led to the well-known confirmation of Einstein's GR in our solar system, are caused by the gravitational field, the curvature of space–time. They are also slightly modified by utterly negligible amounts of DM advected into in the solar system, much smaller than the $10^{-9}$ of Eq. (64), according to our universal key formula. These are the advance of Mercury's perihelion and the twice larger deflection of light rays bypassing the sun's gravitational field.

For the total mass increase of the whole solar system, we consider the limit of the solar system at $D = 2 \text{ ly} = 1.9 \cdot 10^{18}$ cm, almost half the distance to the nearest star Proxima Centauri. From Eq. (46a), we obtain

$$\begin{aligned} M' &= -3\pi\lambda G {g'}^2 Ma + 4\pi\lambda G {g'}^2 MD \\ &\approx 4\pi\lambda G {g'}^2 MD \\ &= M_\odot \left[ 4 \cdot 6.7 \cdot 10^{-8} \pi 1.39 \cdot 10^{-16} 1.9 \cdot 10^{18} \right] \\ &= 2.2 \cdot 10^{-4} M_\odot. \end{aligned} \quad (66)$$

This shows that the total DM fraction maximally accreted by the solar system is also very small, even if we neglect $g_{oo}$ effects.

### 3.5 Dark Matter and the Cosmic Web

Dark matter, identified here with the universe's coherent gravitational field, follows the field equations of GR, Eq. (6), and thus interacts gravitationally also with itself, causing it to clump together, without the opposing pressure present in ordinary matter. This pressure is present in regular matter, arising from the random motion of atoms, molecules, ions and electrons. In quantum gravidynamics (QGD) there could be some "pressure" from gravitons, most of them in localized coherent states, creating the coherent gravidynamic field, the DM. This may set limits to the collapse. Any initially uniform distribution of this DM would have been very unstable to collapse. The formation of the observed cosmic web, DM decorated by regular matter, may have yielded a lower energy configuration. In some nodal places of the cosmic web, where the local total density of all matter and field exceeded the limit for gravitational collapse, this may have yielded the early giant black holes in the first billion years. *There is a tendency to lower the energy of DM,* of the coherent gravitational field of the universe, causing iteratively its own secondary, attractive, gravitational field. This *resulted in its collapse into the skeleton of the cosmic web, early in the cosmic evolution after the big bang*.

One may ask, why the collapse of the DM does not continue leading to one-dimensional filament singularities. The answer may be found in a quantum gravidynamic field theory, that increases the energy when the coherent field is compressed more. Its quanta, introduced for weak field strengths, gravitons, also interact gravitationally. The observed form of the cosmic web could thus help in developing the quantum theory of gravitation.

Practically all of the DM found in galaxies is not locally generated, but is attracted into the galaxy, as we mentioned, being generated at the level of the observable universe. Most of the DM found in clusters of galaxies is also not locally generated, but attracted into the cluster. The attracted DM is distributed proportional to the locally generated coherent gravitational field as we mentioned already in Sect. 2.2. Therefore, the attracted DM avoids the center of mass of galaxies or clusters thereof, where the own generated gravitational field vanishes. It is present usually as a halo, at the periphery, where the own generated field is maximal, and decreases farther out. The fact that DM can be physically independent and separated from ordinary matter, like a mysterious substance, is noticed clearly [9] in the Bullet cluster and agrees in principle with our predictions. As we mentioned before, these important Bullet cluster observations do not disprove our identification of DM, they rather support it strongly.

In conclusion, a new notion of the coherent gravitational field of the universe, described by Eq. (6), with iterative self-interaction, is introduced here and identified as DM. Its behavior is quite different from the electromagnetic field, more like an immaterial substance interacting through gravitational forces and capable to be distributed non-uniformly as "dark matter," lumped together in the cosmic web, attracting and accompanying regular matter.

### 3.6 Connection with the ΛCDM Model

The ΛCDM standard "model" will be modified by our present contribution and will get closer to a simple application of Einstein's GR, based on the results disclosed through our paper with early forms distributed privately to friends and colleagues since 2/10/2022. Our paper differs from a "model." It allows us for the first time to calculate and predict all properties of DM, directly from GR, based on a simple approximation of exact solutions of Einstein's equations. The vast research on DM has not lead to understanding its nature yet [15, 16]. The ΛCDM model encountered several problems in the past, e.g.:

(a) The inability [16] to find known weakly interacting massive particles (WIMPs) [17] that could make up the DM and that would also be "cold," so the structure in the DM is preserved; Absence [15, 16] of any success in searches for axions and sterile neutrinos; The many unsuccessful attempts to find a reasonable explanation of DM;

(b) The inability [15] to explain the observed low number of satellite galaxies [18] that accompany galaxies, calculated to be orders of magnitude higher in DM simulations with hard spheres; Also the peculiar orientation [19] and low mass of the observed satellite galaxies of the Milky Way, Andromeda and others, which all show no resemblance with the existing DM simulations;

(c) The stability and particular configuration of DM patterns in cosmological time intervals [14]. WIMPs would migrate away fast. The coherent gravitational field patterns of the observable universe are stable.

(d) The deviations caused by DM in galactic dynamics depend only on the baryonic (ordinary) matter content, for galaxies of all sizes. This empirical relation, known as the "baryonic Tully–Fisher-relation" finds the total baryonic mass of a galaxy to be proportional to about the fourth power of its maximal rotation speed. *We have derived this empirical relation from our GR coherent gravi-* *tational field theory of DM in* Sect. 3.3. For elliptic galaxies there is a similar Faber-Jackson relation. Why should the deviations caused by DM in galaxies depend only on the baryonic matter and not on DM? As shown in Sect. 2.2 after Eq. (11), in Eqs. (44), (46) and in Sect. 3.3, the essential coherent proportionality of DM patterns with the very small own gravitational field contribution from baryonic galactic matter holds the answer, and also explains the absence of a DM cusp at the center of galaxies, anticipated by ΛCDM simulations. Indeed, as mentioned and shown above, the own gravitational field generated by the galaxy is zero in the center of the galaxy as a general feature and Newtonian physics dominates the whole central area. If the DM advected from the universe is proportional to the field generated by the baryonic matter as shown here, it will never produce a DM maximum in the center of the galaxy, DM will be negligible in the central region as we have shown. We make abstraction of the presence of giant black holes in the center of galaxies, which contribute a few promilles to the galactic mass. We actually found the complete answer to the mentioned exclusive dependence of galactic features on the baryonic matter (and not on DM), in Sects. 3.2 and 3.3: the advected DM itself is dependent on the baryonic matter.

In conclusion, we were able to derive the Tully–Fisher relation and the "key" for calculating the DM concentrations everywhere in lower-scale structures. This allowed us also to finalize the calculation of the rotation curves of stars in galaxies, as shown in Sects. 3.2 and 3.3.

Gravitons, with spin 2, and bosons in general, like to stay "on top of each other," even possibly yielding condensates at low temperatures. The 2.76 °K CMB temperature is in that domain. In general, we expect a better mutual attraction and joint pattern stability of the own galactic field, enhanced by our DM contribution created at the level of the universe, than with the WIMP, axion or neutrino type of standard DM. The GR field equations can lead to a strong amplification effect, an extreme nonlinear enhancement of the locally generated gravitational field. It is equivalent to a "guidance" of the accreted DM distribution by the small locally generated field pattern, according to the key shown in Eq. (59).

All these problems, and many more, can now be treated on the basis of GR. Like in the case of quantum chemistry and experimental chemistry, observation connected with physical, qualitative GR-based, analysis, as shown here, may often be more appropriate than laborious exact solution of EE.

### 3.7 Discussion

Based on GR and on our identification of DM as energy of the coherent gravitational field of the universe, we were able to show in Sects. 2.1–2.6 that both Einstein's equations and, to a certain extent, also their Newtonian approximation, could explain the observed presence of 5 times more DM in the Universe than ordinary, mainly baryonic, matter. Approximative agreement was obtained within an order of magnitude. On this basis we have applied in Sects. 3.1–3.4 our new coherent gravitational field approach based on GR, that had been initiated in Sect. 2.2. It calculated DM successfully, for the first time. Indeed, we were immediately able to derive in Sects. 3.2 and 3.3 the constant velocity profile shown by the galactic rotation curves, from the dominant gg′² terms suggested by the gravitational field self-interaction non-linearity in Einstein's equations, Eq. (6).

Independently, our identification of DM as the coherent gravitational field led us then directly to our derivation in Sect. 3.3 of the Tully–Fisher relation for spiral galaxies. This, in turn, allowed a determination of the coefficient λ in our formula for the resulting "DM key" of the density ρ = λgg′² of DM. This key finally allowed us to prove that the DM-caused velocity plateau we derived in Sect. 3.2 was not disconnected from the known Newtonian velocity profile derived for stars in the central galactic region. The plateau fitted perfectly, smoothly reproducing the empirical curve B in Fig. 1 and verifying our approach.

Moreover, this key also allowed us to prove in Sect. 3.3 that the DM concentration ρ = λg′²GMr/a³ at r = 0 is zero, given a uniform distribution of baryonic matter, constant down to r = 0. This effectively eliminates DM from the near central region, and we were able to calculate the concentration of DM explicitly for the first time everywhere in the galaxy, in agreement with observations. This key should also allow for an exact calculation of the concentration of DM in and around the Bullet cluster with our key for the first time, as we explained at the end of Sect. 3.3. Furthermore, it allowed us also to calculate in Sect. 3.4 the very small DM contribution in our solar system.

It is indeed remarkable that our GR-based identification of DM as the coherent gravitational field of the observable universe, led not only to the classical and GR derivation in cylindric and spherical symmetries, of the observed order of magnitude ratio between DM and ordinary, mainly baryonic, matter in the universe. It also led to the sudden unraveling of the whole DM puzzle.

In essence, in this paper we are predicting, anticipating, or approximating the results of a rigorous, simultaneous, global, all-encompassing *solution* of Einstein's field equations for the universe, for coupled clusters of galaxies and interacting stars in individual galaxies, down to planetary systems like our solar system, coherently including all length scales. A *solution* that agrees with the available evidence.

*Based on the analysis presented here, we could also say that DM does not exist as a separate concept. Indeed, the gravitational field is a feature of space–time in GR, therefore a feature of the event space of our existence. This approach is correct, but impractical, as long as an exact all-encompassing solution of Einstein's equations is not available. The approach used here is a first practical step, that yields a first approximation. As prisoners of the classical, Newtonian point of view, we were misled by the unexpected consequences of Einstein's coupled nonlinear field equations at all length scales, leading to the self-interaction of the gravitational field (across different length scales) and of ordinary matter, as well as to their mutual interaction. This causes forces that mimic the existence of a mysterious, strange, substance, the "Dark Matter." In the way we understand it here, this concept is nevertheless very useful on the way to more exact solutions. It remains to translate our simple GR anticipation results of Sects. 3.1–3.4 into the language of the metric.*

The present analysis can be seen as an initial effort to bridge the gap between the classical physics and GR, by:

1. Actually calculating the amount of DM in the universe, expected even in the lowest order evaluation form Einstein's equations (Eq. 42), which is already double the classical estimate of the gravitational field energy;
2. Deriving the rotation curves in galaxies, Eqs. (45) and (61) with our simple anticipation key of the GR results;
3. Deriving the Tully–Fisher relation Eq. (57) with the same universal formula or key, anticipating the universal solution of Einstein's equations;
4. Deriving the same way the negligible DM amount in the solar system;
5. Deriving DM in principle at all lower scales of the observable universe.

We have shown that even simple, classical-looking calculations, combined with our understanding of the nature of GR, can foster a correct identification and calculation of the DM at all length scales. *The "key" for our approximate solutions is the quantity $\lambda g'^2 = 1.39 \cdot 10^{-16}$ gs$^2$/cm$^4$, which controls the GR accretion of the DM of the universe into lower-scale structures, e.g., clusters, galaxies, planetary systems.* This corresponds to a DM concentration $\rho = \lambda g'^2 g$ being coherently accreted by any gravitational field g generated at lower scale. This accretion comes from the gravitational field strength g′ generated at the level of the observable universe.

The most important feature of Einstein's equations at the universe level is the large creation of DM as gravitational field from DM *which serves mainly as its own source*. This self-generation feature becomes dominant at the level of the universe and leads to the introduction of the mentioned key $\rho = \lambda g'^2 g$. *The two main principles we applied here from GR are this universal-level DM dominance feature and the return to Newton's physics in the local frame in the Euclidian differential neighborhood. This apparent return to classicality even includes the interaction of the locally produced gravitational field with the field generated at the level of the universe, as a form of self-interaction of the global gravitational field, which includes all length scales.*

*In fact, our notion of "coherent accretion of DM generated at the level of the universe to local baryonic-generated gravitational field" admits another interpretation. It is the coherent distribution and inclusion of space–time curvature generated at the universal level to the locally generated, lower-scale curvature. The resulting large-scale curvature is compensated in turn by the expansive pressure effect of the vacuum, that flattens out the universe (see Eq. 6). It causes the universe to be perceived as flat, with no resulting universal curvature. With the curvature compensated at the scale L = 1.000,000 ly of galaxies, we multiply our DM key formula with (L/r)Log(1+r/L) in general, to cover all higher scales of the universe. The physical nature of this vacuum pressure, described by the cosmological constant term $g_{\mu\nu}\Lambda$ in Einsten's equations still needs to be understood.*

Our identification of DM as coherent gravitational field naturally satisfies the following observed cold dark matter properties:

(a) it seems to be non-baryonic, non-leptonic;
(b) it is cold, assumed to be in general equilibrium with the 2.76°K temperature of the CMB;
(c) it is collisionless;
(d) it is dissipationless.
(e) Its presence in galaxies and clusters varies considerably from galaxy to galaxy and among clusters.

In this investigation we have started from a simple, generally accessible, electromagnetic analogy. In electric wires and circuits, through Eq. (1), it divides the electromagnetic field contribution to the mass and energy of the electrons *into a coherent, collective fraction ($\sim N'^2$) and an incoherent, "mechanical" kinetic energy part ($\sim N'$)*. The coherent part comes from the magnetic field energy of the ordered motion of the electrons, while the incoherent part arises from the mechanical kinetic energy of their ordered (drift) motion. *The coherent part is $N'r_e$ times larger than the incoherent part* that includes also any non-electromagnetic contribution. A similar partition $N'r_s$ is introduced in Eq. (2) for the coherent gravitational field energy of the motion of objects, e.g., galaxies in the observable universe, and for the mechanical kinetic energy of the ordered component of their motion. If the average galaxy would be like our Milky way, we would have $N'r_s = 5$, for the observable universe in Eq. (5), but with a more realistic smaller average galaxy we found $N'r_s = 0.35$ in Eq. (8), and left it to the exact GR solution to come closer to the empirical result of 5. In part, this discrepancy may come from differing fractions of DM being included in the mass estimate of the Milky Way and of other galaxies in the universe. And indeed, already in the low order GR approximation of Eq. (42), we got 0.8 $N'r_s$. This is 0.28 with the $N'r_s$ from Eq. (8), and is also twice the classical result 0.4 $N'r_s$ of Eq. (33).

As was mentioned at the end of Sect. 2.1, the use of the estimate $N'r_s$ for the fraction s″ of DM can be justified only for the observable universe as a whole. For larger intermediate distributions of matter and DM along long branches of the cosmic web, it could be used as a first approximation or overestimation, in need of a correcting factor of the order of unity. In lower-scale structures, our key $\lambda gg'^2$ for the DM density should be used instead.

Since DM is identified as coherent gravitational field, it may be quantized *in low-field conditions* in terms of gravitons of spin 2, similar to photons of spin 1 in QED. *This way, we can recognize DM as the gravitational analogue of the cosmic radiation background, CMB*. Indeed, we have now in the cold dark matter the gravitonic counterpart of the photonic CMB. The gravitons could be indirectly observed [2, 3] in the 1/f low-frequency gravidynamic Q1/fE fluctuation spectra of macroscopic matter flows, just like e.l.f. photons are observed in the same way in the technically important electrodynamic Q1/fE case [20–22]; see the supplementary information referenced below.

This QED quantum 1/f noise theory was right away translated into the gravidynamic domain in 1975 by Handel and the relation between the two gravitational Q1/fE (coherent and conventional) was established by him in 2003. To find this relation, he had to evaluate the ratio [Eq. (2)] between the "coherent- field" and "incoherent/mechanical" gravitational energy, both referring to the ordered flow motion. This led directly to our present paper that identifies and calculates DM everywhere, as an anticipation of numerical or analytical future GR calculations.

Our present application of GR differs both from the standard DM model that keeps looking for hidden DM particles, or from the Modified Newtonian Dynamics (MOND).

MOND [23–25] modifies Newton's law (and implicitly GR) at low gravitational fields, to explain the rotation curves of galaxies and other DM phenomena. MOND has no clear theoretical basis, while our approach is on the theoretical ground of GR and of its local Newtonian approximation. It introduces only one new parameter, λ, fixed by the empirical constant in the baryonic Tully–Fisher relation. This parameter can be derived in principle from Einstein's field equations. Its derivation would be performed by descending from the largest scale, where the gravitational field dominates as its own main source in GR.

## Appendix 1

At this point we briefly outline the elementary derivation of the field energy of the columnar universe, and of the number 0.25 added to the logarithm in Eqs. (1) and (2) in *cylindrical symmetry*, for uniform electric current j and charge ρ distributions. This is done first in the rest frame, in which the column moves, and then also in the co-moving (local) frame, in which the column is at rest:

(1) Rest frame (of embedding space)

$$\text{Ampere: } 2\pi r B = 4\pi^2 r^2 j/c; \ B = 2\pi r j/c \ (r < a)$$
$$W_{r<a} = \int_{r<a} \mathbf{B}^2 d\tau / 8\pi = (\pi j/c)^2 \int r^3 dr = J^2/4c^2 \quad (67)$$

$$r > a: \mathbf{B} = 2J/rc; \quad W_{r>a} = \int B^2 d\tau / 8\pi$$
$$= (J^2/c^2) \int dr/r = (J^2/c^2) \ln(R_0/a); J = \pi a^2 j = N'ev. \quad (68)$$

$$W_{Ktotal} = (J^2/c^2)[0.25 + \ln(R_0/a)],$$
$$W_{Ktotal} = W_0 v^2/c^2. \quad (69)$$

(2) Co-moving (local) frame of the column

$$\text{Gauss}: \ 4\pi \cdot \pi\rho r^2 = 2\pi r E \quad (r < a)$$

$$\begin{aligned} E_{r<a} = 2\pi r\rho; \quad W_{r<a} &= \int_{r<a} \mathbf{E}^2 d\tau/8\pi \\ &= (\pi\rho)^2 \int r^3 dr = Q'^2/4; \quad Q' = \pi r^2 \rho = N'e. \end{aligned} \tag{70}$$

$$E_{r>a} = 2Q'/r; \quad W_{r>a} = \int_{r>a} \mathbf{E}^2 2\pi r dr/8\pi = Q'^2 \ln(R_0/a). \tag{71}$$

$$W_{Q' \text{total}} = Q'^2 [0.25 + \ln(R_0/a)] = W_0 \tag{72}$$

This is the energy in the co-moving frame, going with the column, where the energy is only electrostatic. In the rest frame, in which the column has velocity v, the components of the field's energy–momentum vector are: the time-like energy component

$$\begin{aligned} W = \gamma W_0 &= \gamma W_{Q' \text{ total}} = \gamma Q'^2 [0.25 + \ln(R_0/a)] \\ &= W/\gamma + (\gamma - 1)W/\gamma = W_{Q' \text{ total}} + W_{K \text{ total}}, \end{aligned} \tag{73}$$

and the space-like component

$$\gamma W_{Q' \text{ total}} v/c = \gamma Q'^2 [0.25 + \ln(R_0/a)] v/c. \tag{74}$$

Here $\gamma = (1 - v^2/c^2)^{-1/2} \approx 1 + v^2/2c^2$).

The last 2 forms of Eq. (73) correspond to the sum of Eq. (69) and (72). In this sum, the total $v^2/c^2$ contribution is shown explicitly with a factor ½, like in a non-relativistic kinetic energy.

The coherent field energy terms, both the kinetic $W_{Ktotal}$ part due to current J, and static $W_{Q' \text{ total}}$ from charge per unit length Q′, are thus proportional to $N'^2$. ***The rest mass corresponding to the field energy in the co-moving system is $W_0/c^2$. The same rest mass result is obtained in the embedding rest frame.***

There is a nice similarity with the gravitational case in cylindrical symmetry, as shown in Sects. 2.1, 2.3, 2.5 and 2.6.

## Appendix II

Here we use the Schwarzschild solution of Einstein's Equations and Eqs. (13)–(15) to calculate $t_{00}$, which was used in Eq. (38), with $\eta_{\mu\nu}$ defined at Eq. (12),

$$8\pi G t_{00} = (1/2)\left\{-h_{00}R^{(1)\lambda}_{\lambda} + h^{\rho\sigma}R^{(1)}_{\rho\sigma} + R^{(2)}_{00} - \eta^{\rho\sigma}R^{(2)}_{\rho\sigma}\right\} + O(h^3). \tag{75}$$

The first two terms in curly brackets do not contribute up to second order in h, as we see below in Eqs. (77).

According to Eq. (35), in lowest (second) order in $h_{\mu\nu}$, we set $h_{11}=h_{22}=h_{33}=-4A-6A^2$; $h_{00}=-4A+8A^2$, with $A=mN/2r$. This yields, with $\nabla^2(1/r)=0$ ($r>a$) and with $\partial/\partial x^0 = 0$, from Eq. (14) to first order in $h_{\mu\nu}$ and lowest order in A

$$R^{(1)}_{00} = 2m^2N^2/r^4; \tag{76}$$

it yields $O(h^3)$ with the $h_{\mu\nu}$ factor in front and will be neglected in this lowest approximation. In a similar way we can write

$$R^{(1)}_{00} = R^{(1)}_{11} = R^{(1)}_{22} = R^{(1)}_{33} = 0 \tag{77}$$

$$R^{(1)}_{10} = R^{(1)}_{20} = R^{(1)}_{30} = R^{(1)}_{01} = R^{(1)}_{02} = R^{(1)}_{03} = 0$$

$$R^{(1)}_{12} = R^{(1)}_{23} = R^{(1)}_{31} = R^{(1)}_{21} = R^{(1)}_{32} = R^{(1)}_{13} = 0$$

These are actually not zero, but yield $O(h^3)$ with the factor $h_{\mu\nu}$ in front of them.

We also obtain from Eq. (15) to second order in $h_{\mu\nu}$, and lowest order in A, with $h_{00}=h_{11}=h_{22}=h_{33}=-4A=h$, $\nabla^2 h=0$ and $h_{\mu\nu}=0$ for $\mu \neq \nu$:

$$\begin{aligned} R^{(2)}_{00} &= (h/2)[\nabla^2 h] + (1/4)[2\partial h^\nu_\sigma/\partial x^\nu - \partial h^\nu_\nu/\partial x^\sigma][-\partial h_{00}/\partial x_\sigma] \\ &\quad - (1/4)[\partial h_{\sigma 0}/\partial x^\lambda - \partial h_{\lambda 0}/\partial x^\sigma][\partial h^\sigma_0/\partial x_\lambda - \partial h^\lambda_0/\partial x_\sigma] \end{aligned} \tag{78}$$

This yields

$$\begin{aligned} R^{(2)}_{00} &= (1/4)[2\partial h^\nu_\sigma/\partial x^\nu - \partial h^\nu_\nu/\partial x^\sigma][-\partial h_{00}/\partial x_\sigma] - (1/4)[\partial h_{\sigma 0}/\partial x^\lambda - \partial h_{\lambda 0}/\partial x^\sigma][\partial h^\sigma_0/\partial x_\lambda - \partial h^\lambda_0/\partial x_\sigma]. \\ &= (1/4)[2(\nabla h)^2 + 2(\nabla h)^2] - (1/4)[-(\nabla h)^2 - (\nabla h)^2] \\ &= (3/2)(\nabla h)^2 = 6m^2N^2/r^4. \end{aligned} \tag{79}$$

We also obtain from Eq. (15)

$$\begin{aligned} R^{(2)}_{11} &= (1/2)h^{\lambda\nu}[\partial^2 h_{\lambda\nu}/\partial x^1 \partial x^1 - \partial^2 h_{1\nu}/\partial x^1 \partial x^\lambda - \partial^2 h_{\lambda 1}/\partial x^\nu \partial x^1 + \partial^2 h_{11}/\partial x^\nu \partial x^\lambda] \\ &\quad + (1/4)[2\partial h^\nu_\sigma/\partial x^\nu - \partial h^\nu_\nu/\partial x^\sigma][\partial h^\sigma_1/\partial x^1 + \partial h^\sigma_1/\partial x^1 - \partial h_{11}/\partial x_\sigma] \\ &\quad - (1/4)[\partial h_{\sigma 1}/\partial x^\lambda + \partial h_{\sigma\lambda}/\partial x^1 - \partial h_{\lambda 1}/\partial x^\sigma][\partial h^\sigma_1/\partial x_\lambda + \partial h^{\sigma\lambda}/\partial x^1 - \partial h^\lambda_1/\partial x_\sigma] \end{aligned} \tag{80}$$

$$\begin{aligned} &= (1/2)[h\partial^2 h/\partial x^{12} + h\nabla^2(h)] \\ &\quad + (1/4)[2(\partial h/\partial x^1)^2 + 2(\partial h/\partial x^1)^2 - 2(\nabla h)^2 - 2(\partial h/\partial x^1)^2 - 2(\partial h/\partial x^1)^2 + 2(\nabla h)^2] \\ &\quad - (1/4)[(\nabla h)^2 + (\partial h/\partial x^1)^2 + (\partial h/\partial x^1)^2 - (\partial h/\partial x^1)^2 + 4(\partial h/\partial x^1)^2 \\ &\quad - (\partial h/\partial x^1)^2 - (\partial h/\partial x^1)^2 - (\partial h/\partial x^1)^2 + \nabla h)^2]. \end{aligned} \tag{81}$$

We obtain thus

$$R^{(2)}_{11} = (1/2)[h\partial^2 h/\partial x^{12} - (\nabla h)^2 + (\partial h/\partial x^1)^2];$$
$$R^{(2)}_{22} = (1/2)[h\partial^2 h/\partial x^{22} - (\nabla h)^2 + (\partial h/\partial x^2)^2]; \quad (82)$$
$$R^{(2)}_{33} = (1/2)[h\partial^2 h/\partial x^{32} - (\nabla h)^2 + (\partial h/\partial x^3)^2];$$

From Eq. (75) we get with (77) and $\nabla^2 h = 0$

$$8\pi G t_{00} = (1/2)\left\{R^{(2)}_{00} - \left[R^{(2)}_{00} - R^{(2)}_{11} - R^{(2)}_{22} - R^{(2)}_{33}\right]\right\} + O(h^3) \approx -(1/2)(\nabla h)^2 = -2m^2 N^2/r^4. \quad (83)$$

This result $t_{00} = 2Gm^2N^2/8\pi r^4 = mc^2 N^2 r_s/8\pi r^4$ was inserted above in Eqs. (38) and (39) of Sect. 2.6.

**Supplementary Information** The online version contains supplementary material available at https://doi.org/10.1007/s10701-023-00705-x.

**Author contributions** This manuscript is a teamwork. Both authors have verified the submitted paper.

**Declarations**

**Conflict of interest** The authors declare no competing interests.

**Open Access** This article is licensed under a Creative Commons Attribution 4.0 International License, which permits use, sharing, adaptation, distribution and reproduction in any medium or format, as long as you give appropriate credit to the original author(s) and the source, provide a link to the Creative Commons licence, and indicate if changes were made. The images or other third party material in this article are included in the article's Creative Commons licence, unless indicated otherwise in a credit line to the material. If material is not included in the article's Creative Commons licence and your intended use is not permitted by statutory regulation or exceeds the permitted use, you will need to obtain permission directly from the copyright holder. To view a copy of this licence, visit http://creativecommons.org/licenses/by/4.0/.